\documentclass[a4paper,11pt]{article}
\pdfoutput=1 

\usepackage{jcappub} 
\usepackage[T1]{fontenc} 
\usepackage{xfrac}
\usepackage{amsmath}
\usepackage{mathtools}
\usepackage{enumitem}
\usepackage{color}
\usepackage{aas_macros}

\usepackage[T1]{fontenc} 
\usepackage{natbib}
\definecolor{internationalkleinblue}{rgb}{0.0, 0.18, 0.65}
\hypersetup{colorlinks=true, urlcolor=internationalkleinblue, linkcolor=internationalkleinblue, citecolor=internationalkleinblue}

\title{Breaking baryon-cosmology degeneracy with the electron density power spectrum}

\author[a]{Andrina Nicola}
\author[a,b]{Francisco Villaescusa-Navarro}
\author[a,b]{David N. Spergel}
\author[a,c]{Jo Dunkley}
\author[d,b]{Daniel Angl\'es-Alc\'azar}
\author[e,f,g]{Romeel Dav\'e}
\author[b,h]{Shy Genel}
\author[i]{Lars Hernquist}
\author[j]{Daisuke Nagai}
\author[b]{Rachel S. Somerville}
\author[k,b]{Benjamin D. Wandelt}

\affiliation[a]{Department of Astrophysical Sciences, Princeton University, Peyton Hall, Princeton, NJ 08544, USA}
\affiliation[b]{Center for Computational Astrophysics, Flatiron Institute, 162 Fifth Avenue, New York, NY 10010, USA}
\affiliation[c]{Department of Physics, Princeton University, Princeton, NJ 08544, USA}
\affiliation[d]{Department of Physics, University of Connecticut, 196 Auditorium Road, Storrs, CT 06269, USA}
\affiliation[e]{Institute for Astronomy, University of Edinburgh, Royal Observatory, Edinburgh EH9 3HJ, UK}
\affiliation[f]{Department of Physics \& Astronomy, University of the Western Cape, Cape Town 7535, South Africa}
\affiliation[g]{South African Astronomical Observatories, Observatory, Cape Town 7925, South Africa}
\affiliation[h]{Columbia Astrophysics Laboratory, Columbia University, New York, NY 10027, USA}
\affiliation[i]{Center for Astrophysics $|$ Harvard \& Smithsonian, 60 Garden St, Cambridge, MA 01238, USA}
\affiliation[j]{Department of Physics, Yale University, New Haven, CT 06520, USA}
\affiliation[k]{Institut d'Astrophysique de Paris (IAP), UMR 7095, CNRS, Sorbonne Universit\'e, France}

\emailAdd{anicola@astro.princeton.edu}

\abstract{Uncertain feedback processes in galaxies affect the distribution of matter, currently limiting the power of weak lensing surveys. 
If we can identify cosmological statistics that are robust against these uncertainties, or constrain these effects by other means, then we can enhance the power of current and upcoming observations from weak lensing surveys such as DES, Euclid, the Rubin Observatory, and the Roman Space Telescope. In this work, we investigate the potential of the electron density auto-power spectrum as a robust probe of cosmology and baryonic feedback. We use a suite of (magneto-)hydrodynamic simulations from the CAMELS project and perform an idealized analysis to forecast statistical uncertainties on a limited set of cosmological and physically-motivated astrophysical parameters. We find that the electron number density auto-correlation, measurable through either kinematic Sunyaev-Zel'dovich observations or through Fast Radio Burst dispersion measures, provides tight constraints on $\Omega_{m}$ and the mean baryon fraction in intermediate-mass halos, $\bar{f}_{\mathrm{bar}}$. By obtaining an empirical measure for the associated systematic uncertainties, we find these constraints to be largely robust to differences in baryonic feedback models implemented in hydrodynamic simulations. We further discuss the main caveats associated with our analysis, and point out possible directions for future work.}

\begin{document}
\maketitle
\flushbottom

\section{Introduction} \label{sec:intro}

Observational cosmology is currently undergoing a transformational change, through programs that will deliver a wealth of astronomical data. In the optical wavelength range, these include the Dark Energy Survey (DES)\footnote{\url{https://www.darkenergysurvey.org/}.}, the Hyper-Suprime Cam Survey (HSC)\footnote{\url{https://hsc.mtk.nao.ac.jp/ssp/0}.}, the Rubin Observatory Legacy Survey of Space and Time (LSST)\footnote{\url{https://www.lsst.org/}.}, Euclid\footnote{\url{https://www.euclid-ec.org/}.}, and the Roman Space Telescope\footnote{\url{https://roman.gsfc.nasa.gov/}.}. These surveys will enable high-precision measurements of galaxy clustering, weak gravitational lensing (WL) and their cross-correlation down to small spatial scales. The statistical properties of cosmic fields on small spatial scales carry a wealth of cosmological and astrophysical information, and it will be crucial to extract this information in order to realize the full potential of both current and future surveys (for observational studies, see e.g. Refs.~\cite{Huang:2021, Hadzhiyska:2021}).  

In weak gravitational lensing analyses, the largest theoretical uncertainties stem from an incomplete knowledge of baryonic feedback processes and their impact on the matter distribution. The distribution and properties of baryons in the Universe are determined by a multitude of complex, nonlinear effects spanning a wide range of scales, such as gas cooling, star formation, and feedback due to Active Galactic Nuclei (AGN) and Supernovae (SNe) (see e.g. Refs.~\cite{Somerville:2015, Naab:2017, Vogelsberger:2020}). These processes affect the matter power spectrum in different ways: on the one hand, gas cooling enhances the power spectrum at small spatial scales, as it allows gas to cluster on smaller scales than the Dark Matter. Feedback due to AGN and SNe on the other hand, smoothes the matter distribution and thus causes a reduction in clustering power on intermediate and small scales (see e.g. Ref.~\cite{vanDaalen:2011}). The interplay of these processes is difficult to model analytically, and therefore theoretical predictions for the properties of baryons and their consequences for the underlying matter distribution often rely on hydrodynamic simulations. Examples of studies of the baryon distribution using hydrodynamic simulations include Refs.~\cite{Chisari:2018, vanDaalen:2011, vanDaalen:2020}, while analytic or semi-analytic studies are presented in Refs.~\cite{Mead:2015, Mead:2020, Mead:2021, Schneider:2015, Schneider:2019, Giri:2021}, amongst others. While current simulations can accurately model gravitational interactions and capture the effects of gas cooling, the enormous range of physical scales associated with galaxy formation require that all numerical simulations use sub-grid models to approximate the effects of star formation, stellar feedback, massive black hole growth, and AGN feedback (see e.g. Refs.~\cite{Genel:2014, Somerville:2015, Angles-Alcazar:2017}). 

Differences in sub-grid physics implementations lead to significantly different predictions for the effects of baryons as obtained from these simulations (see e.g. Ref.~\cite{Chisari:2018, vanDaalen:2020, Villaescusa:2020a}). CAMELS is a large suite of (magneto-)hydrodynamic simulations, created to improve our understanding of these theoretical uncertainties \cite{Villaescusa:2020a}. The suite comprises more than 2,000 state-of-the-art hydrodynamic simulations spanning a wide range of cosmological and baryonic feedback parameters. The simulations also incorporate two distinct sub-grid models, following those of the IllustrisTNG \cite{Weinberger:2017, Pillepich:2018} and SIMBA \cite{Dave:2019} simulations. As shown in Ref.~\cite{Villaescusa:2020a}, these differences primarily lead to higher gas temperatures in the intergalactic medium and stronger baryonic effects on the matter power spectrum in the SIMBA simulations as compared to IllustrisTNG, since the former incorporate feedback in the form of fast AGN jets that propagate to large spatial scales \cite{Dave:2019, Borrow:2020, Christiansen:2020}. Due to computational constraints, the CAMELS simulations cover only a relatively small volume of $(25 \; h^{-1} \; \mathrm{Mpc})^{3}$ each, but offer a uniquely useful data set for investigating effects of baryons on cosmic fields and related uncertainties due to implementation differences in simulations.

The distribution and thermodynamic properties of gas in the Universe can be constrained using several observables, such as X-ray emission, secondary Cosmic Microwave Background (CMB) anisotropies due to the kinematic and the thermal Sunyaev-Zel'dovich effects (kSZ, tSZ) \cite{Sunyaev:1972}, 21cm emission \cite{Pritchard:2012}, and Fast Radio Bursts (FRB). FRBs are short, luminous radio pulses of extragalactic origin \cite{Petroff:2019} and the distribution of their observed dispersion measures (DM) is emerging as a promising tool to probe ionized electrons in the Universe (see e.g. Refs.~\cite{McQuinn:2014, Masui:2015, Fujita:2017, Munoz:2018, Ravi:2019, Macquart:2020, Shirasaki:2021}). A recent analysis of CHIME\footnote{\url{https://chime-experiment.ca/en.}} data reported the first detection of a cross-correlation between the angular distribution of FRBs with several galaxy catalogs \cite{Rafiei-Ravandi:2021}, and used these measurements to derive constraints on the redshift distribution of FRB hosts. 

In this work, we take a step towards investigating the potential of these baryonic observables to constrain cosmological and baryon feedback parameters. Specifically, we focus on the three-dimensional electron number density power spectrum, $P_{ee}(k)$\footnote{For all power spectra, $k$ denotes the modulus of the three-dimensional wave vector.}, which underlies observations of the kSZ and FRBs. The constraints obtained in these analyses can be used to tune hydrodynamic simulations, thereby improving theoretical models for the suppression of the matter power spectrum and thus the weak lensing signal, which will be crucial for optimally benefiting from current and future weak lensing data. We use the CAMELS suite of hydrodynamic simulations to quantify the constraining power of $P_{ee}(k)$ on a limited set of cosmological and astrophysical parameters, robust to systematic uncertainties in baryonic physics: Specifically, we use a simplified analysis to forecast statistical uncertainties on $\Omega_{m}$, $\sigma_{8}$, and the mean baryon fraction in intermediate-mass halos, $\bar{f}_{\mathrm{bar}}$. Furthermore, we estimate systematic uncertainties associated with these analyses and thus assess their robustness to uncertainties in baryonic feedback models. Our results suggest that the electron power spectrum is promising for obtaining constraints on both cosmology and astrophysics, and that the baryon fraction in intermediate-mass halos appears to be a fairly accurate predictor for baryonic feedback strength (see also Refs.~\cite{vanDaalen:2020, Giri:2021, Shirasaki:2021}). We describe the major caveats associated with this analysis and outline possible avenues for future work.

This paper is organized as follows. In Sec.~\ref{sec:methods}, we give an overview of the cosmological quantities considered in our analysis, the simulations used, and the methods employed. We present our results in Sec.~\ref{sec:results}. In Sec.~\ref{sec:caveats}, we discuss the caveats associated with our analysis and we conclude in Sec.~\ref{sec:conclusions}. Implementation details are deferred to the Appendices. 

\section{Methods} \label{sec:methods}

\subsection{Theoretical quantities underlying observational probes of the baryon distribution}\label{ssec:electron-x-corrs}

The cosmological statistics underlying auto- and cross-correlations between observational probes of the baryon distribution are the three-dimensional auto- and cross-power spectra of the matter and galaxy distributions with the electron distribution as well as their pressure. Examples include the electron number density power spectrum, $P_{ee}(k, z)$, the matter-electron cross-power spectrum, $P_{me}(k, z)$, the galaxy-electron power spectrum, $P_{ge}(k, z)$, or the cross-power spectrum between the electron density and pressure distribution, $P_{pe}(k, z)$. The quantities involving the electron distribution can be related to the matter distribution through a bias factor (see e.g. Ref.~\cite{Takahashi:2021}), which has been extensively studied in the literature \cite{vanDaalen:2011, Chisari:2018, vanDaalen:2020}. Despite the fact that the three-dimensional power spectra introduced above are not directly observable with current surveys, in this work we perform an idealized analysis and use a suite of hydrodynamic simulations to obtain forecasted constraints on cosmological and astrophysical parameters, focusing particularly on $P_{ee}(k, z)$. While we also looked at all the other three-dimensional power spectra discussed above, we found the results not to be as promising as those obtained for $P_{ee}(k, z)$. In the remainder of this work, we therefore focus on the latter quantity, but describe the results for the other power spectra when useful. We additionally defer an analysis of the cosmological observables associated to these power spectra to future work, but summarize the generic relations between power spectra and observables in Appendix \ref{ap:sec:P2obs}. 

\subsection{Simulations} \label{ssec:sims}

The CAMELS suite is a set of 4,233 cosmological simulations with volumes of $(25 \; h^{-1} \; \mathrm{Mpc})^{3}$ each, consisting of 2,049 N-body and 2,184 hydrodynamic simulations \cite{Villaescusa:2020a}. All N-body simulations in this suite have a Dark Matter mass resolution of $7.75\times 10^{7} \;\Omega_{m}/0.3 \; h^{-1} M_{\odot}$\footnote{Here, $M_{\odot}$ denotes the mass of the Sun.}, while the Dark Matter and gas mass resolutions of the hydrodynamic simulations are $6.49\times 10^{7} \;(\Omega_{m}-\Omega_{b})/0.251 \; h^{-1} M_{\odot}$ and $1.27 \times 10^{7} \; h^{-1} M_{\odot}$, respectively. The magneto-hydrodynamic simulations have been run using two different codes, AREPO\footnote{\url{https://arepo-code.org}.} \cite{Springel:2010, Weinberger:2020} and GIZMO\footnote{\url{http://www.tapir.caltech.edu/~phopkins/Site/GIZMO.html}.} \cite{Hopkins:2015}, and include galaxy formation modeling following IllustrisTNG \cite{Weinberger:2017, Pillepich:2018} and SIMBA \cite{Dave:2019}, respectively. Unless noted otherwise, all results in this work have been derived using the IllustrisTNG suite. The simulations are run varying a set of six different cosmological and astrophysical parameters, $\boldsymbol{\vartheta} = (\Omega_{m}, \sigma_{8}, A_{\mathrm{SN}1}, A_{\mathrm{SN}2}, A_{\mathrm{AGN}1}, A_{\mathrm{AGN}2})$, where $\Omega_{m}$ is the fractional matter density today, $\sigma_{8}$ is the r.m.s. of linear matter fluctuations in spheres of comoving radius 8 $h^{-1}$ Mpc at $z=0$ and $A_{\mathrm{SN}1}, A_{\mathrm{SN}2}, A_{\mathrm{AGN}1}, A_{\mathrm{AGN}2}$ are astrophysical parameters controlling the strength of feedback due to SNe and AGN in the simulations. Specifically for IllustrisTNG, $A_{\mathrm{SN}1}$ parameterizes the galactic wind energy per unit star formation rate (SFR), $A_{\mathrm{SN}2}$ sets the galactic wind speed, $A_{\mathrm{AGN}1}$ parametrizes the energy released in kinetic mode AGN feedback per unit Black Hole (BH) accretion rate, and finally $A_{\mathrm{AGN}2}$ controls the burstiness and ejection speed of kinetic mode AGN feedback. For SIMBA on the other hand, $A_{\mathrm{SN}1}$ controls the normalization of the mass loading factor for stellar feedback, $A_{\mathrm{SN}2}$ parametrizes the galactic wind speed, $A_{\mathrm{AGN}1}$ sets the total momentum flux ejected in both radiative and jet mode AGN feedback, and $A_{\mathrm{AGN}2}$ determines the outflow speed in jet mode AGN feedback. These parameters are allowed to vary within wide prior ranges  $\Omega_{m} \in [0.1, 0.5]$, $\sigma_{8} \in [0.6, 1.0]$, $A_{\mathrm{SN}1}, A_{\mathrm{AGN}1} \in [0.25, 4.0]$, $A_{\mathrm{SN}2}, A_{\mathrm{AGN}2} \in [0.5, 2.0]$, and the fiducial model is specified by $\Omega_{m} = 0.3$, $\sigma_{8} = 0.8$ and $A_{\mathrm{SN}1} = A_{\mathrm{AGN}1} = A_{\mathrm{SN}2} = A_{\mathrm{AGN}2} = 1$\footnote{The remaining cosmological parameters are fixed to $\Omega_{b} = 0.049, \; h = 0.6711, \; n_{s} = 0.9624$ within $\Lambda$CDM.}.

Both suites of hydrodynamic simulations are subdivided into four sets, and the three sets used in this work are summarized below:
\begin{itemize}[label={}]
\item \textit{Latin Hypercube (LH)}: The \textit{LH} set consists of 1,000 simulations, run on a latin hypercube spanning the six cosmological and astrophysical parameters varied in CAMELS over their full prior ranges. Each realization is run with different initial conditions. This set is designed to make predictions over the full parameter space spanned by the simulations.
\item \textit{1-Parameter (1P)}: The \textit{1P} set consists of 61 simulations, run varying a single simulation parameter at the time over a regular grid in the prior range. In contrast to the \textit{LH} simulations, all \textit{1P} realizations share the same initial conditions. This set of simulations is designed to explore the effects of single parameter changes on simulation observables.
\item \textit{Cosmic Variance (CV)}: The \textit{CV} set consists of 27 simulations, run for a fiducial cosmological and astrophysical model but for varying random phases in the initial conditions. These simulations are designed to explore the effects of cosmic variance on simulation observables\footnote{Throughout this manuscript, we follow the convention used in Ref.~\cite{Villaescusa:2020a} and employ the term cosmic variance as a synonym for sample variance associated to the choice of simulated or observed volume, rather than the size of our physical Universe.}.
\end{itemize}
For further details regarding the CAMELS simulations, the reader is referred to Ref.~\cite{Villaescusa:2020a} and references therein.

\subsection{Power spectrum measurement}

To investigate the potential of the electron power spectrum to jointly constrain cosmology and astrophysics, we measure the electron overdensity field, $\delta_{e}$, directly from the simulation snapshots\footnote{We compute the number of electrons in each gas cell using its volume, density and \texttt{ElectronAbundance} (see \url{https://www.tng-project.org/data/docs/specifications/}).}. For completeness, we note that we additionally compute the galaxy overdensity field $\delta_{g}$, the matter overdensity field $\delta_{m}$, and the mass-weighted pressure field $p$, as described in Appendix \ref{ap:sec:comp_fields}.

For each simulation considered in our analysis, we then compute three-dimensional power spectra, focusing on $P_{ee}(k)$, which constitutes the underlying statistic probed in auto-correlation analyses of FRB and kSZ observations.

We discretize all cosmological fields into $512^{3}$ voxels and compute power spectra by averaging over wave vector $k$-bins with a width equal to the fundamental frequency, $k_{F} = 2\pi/L$, where $L$ denotes the length of the simulation box, $L = 25~h^{-1} \mathrm{Mpc}$. For all power spectrum calculations, we employ the publicly-available \texttt{Pylians} library \cite{Villaescusa:2018}\footnote{The code can be found at \url{https://github.com/franciscovillaescusa/Pylians3}.}.

\section{Results}\label{sec:results}

Before using CAMELS to forecast constraints on cosmological and astrophysical parameters from $P_{ee}(k)$ in Sec.~\ref{ssec:constr_fdbck}, we assess the sensitivity of this statistic to variations in baryonic feedback strength. We particularly aim to identify the main physical processes driving the power spectrum responses within CAMELS. In Sec.~\ref{ssec:fbar_preds}, we additionally investigate observable, physical quantities that allow us to predict the strength of baryonic feedback in the simulations.

In the following, we focus on the electron auto-power spectrum $P_{ee}(k)$ at redshift $z=0$, and only discuss other quantities and redshifts where needed.

\subsection{Sensitivity to baryonic feedback effects}\label{ssec:fdbck_effects}

As described in Sec.~\ref{ssec:sims} and Ref.~\cite{Villaescusa:2020a}, the strength of baryonic feedback in each CAMELS simulation is parameterized by four parameters, $A_{\mathrm{SN}1}, A_{\mathrm{SN}2}, A_{\mathrm{AGN}1}, A_{\mathrm{AGN}2}$. In order to assess the impact of the different feedback processes on the power spectra considered in this work, we focus on the \textit{1P} simulation set, as it allows us to isolate the responses to each parameter. 

\begin{figure*}
\begin{center}
\includegraphics[width=0.49\textwidth]{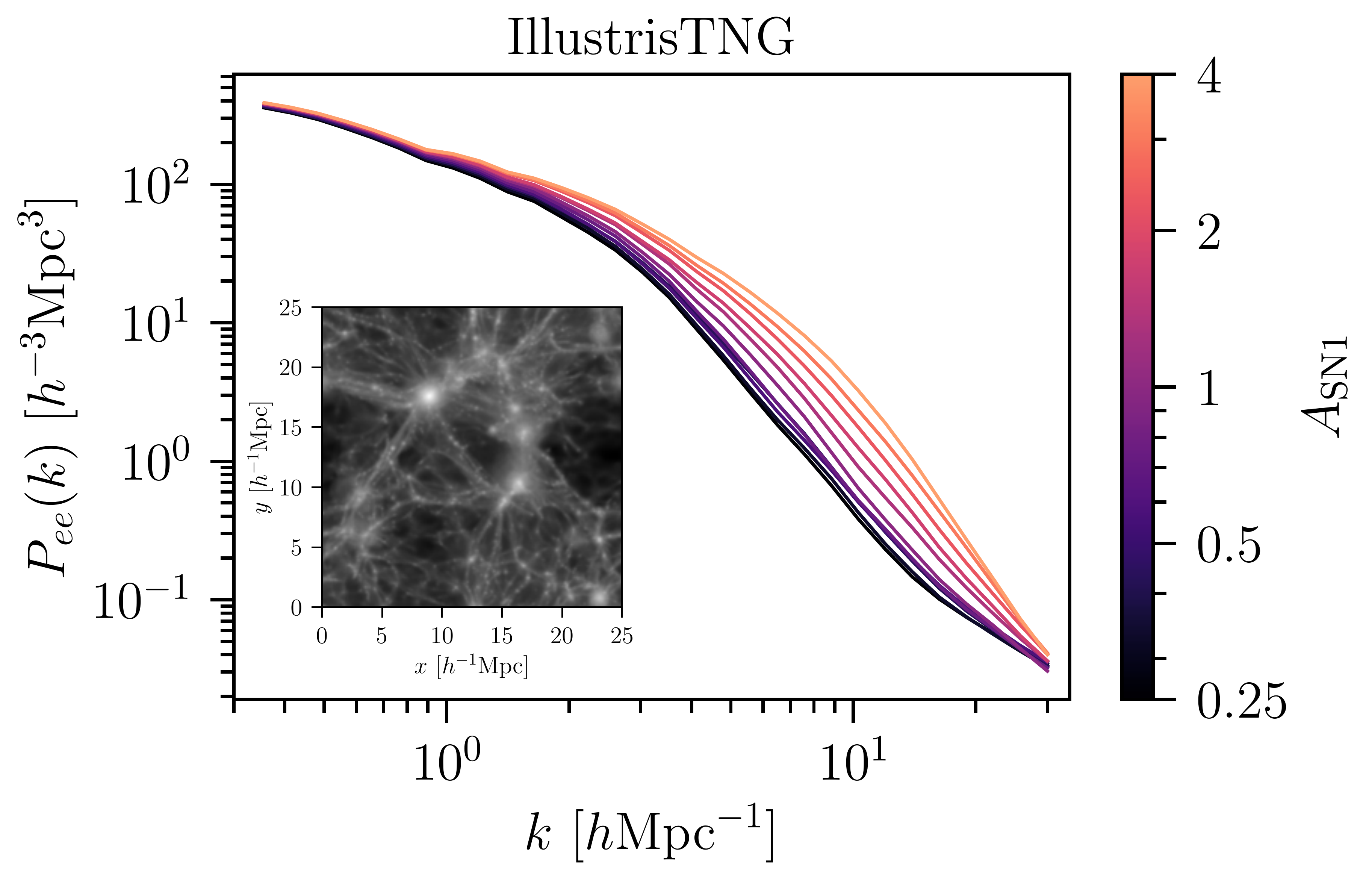}
\includegraphics[width=0.49\textwidth]{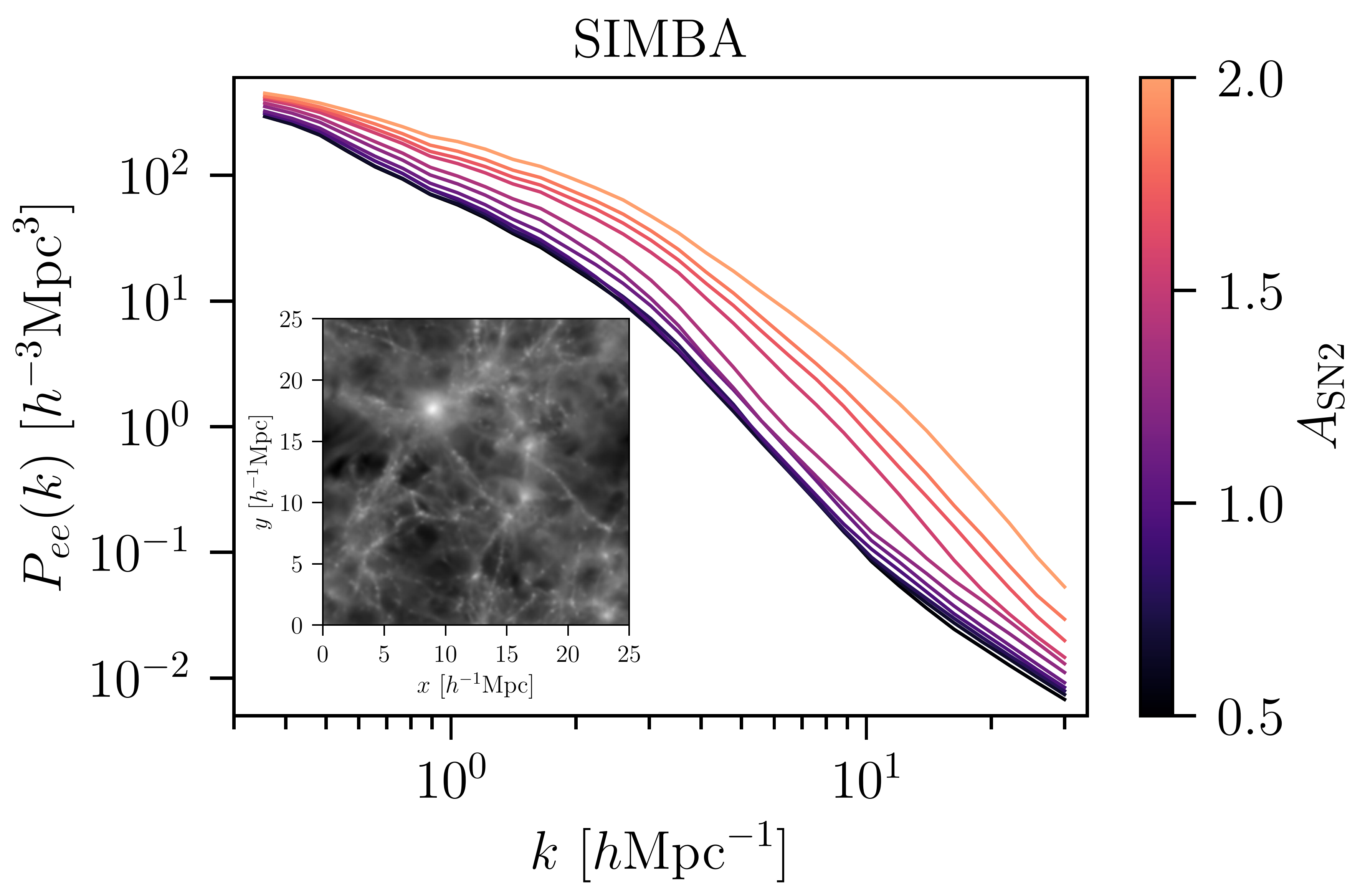}
 \caption{Electron density auto-power spectra as measured from the \textit{1P} set of simulations in IllustrisTNG and SIMBA as a function of the CAMELS astrophysical parameter with the most significant effect, i.e. $A_{\mathrm{SN}1}$ for IllustrisTNG, and $A_{\mathrm{SN}2}$ for SIMBA. For both panels, the insets show a two-dimensional image of the electron number density obtained for the fiducial model (corresponding to $A_{\mathrm{SN}1} = A_{\mathrm{SN}2} = 1$) in each simulation with matched initial conditions.}
\label{fig:pkee_z0_1P_IllustrisTNG_SIMBA}
\end{center}
\end{figure*}

We first compare the electron power spectra measured in the two simulations suites, IllustrisTNG and SIMBA, and find that $P_{ee}(k)$ from IllustrisTNG responds most strongly to the amount of energy injected by wind-driven SNe feedback through $A_{\mathrm{SN}1}$, while the power spectra in SIMBA are most sensitive to galactic wind speed through $A_{\mathrm{SN}2}$ and outflow speed  in jet mode AGN feedback through $A_{\mathrm{AGN}2}$. This means that the sub-grid physics models implemented in IllustrisTNG and SIMBA lead to significantly different physical processes driving baryonic feedback effects in the two simulations. In Fig.~\ref{fig:pkee_z0_1P_IllustrisTNG_SIMBA}, we show the power spectrum responses with respect to $A_{\mathrm{SN}1}$ for IllustrisTNG and $A_{\mathrm{SN}2}$ for SIMBA, as evaluated for the 1P sets\footnote{We do not show results for $A_{\mathrm{AGN}2}$ for SIMBA, as the trends are similar to those shown for $A_{\mathrm{SN}2}$.}. As can be seen, baryonic feedback affects the power spectra in SIMBA on larger scales than it does in IllustrisTNG, which might lead to the more diffuse electron distribution observed in SIMBA and shown in the insets (see also Refs.~\cite{Borrow:2020, Christiansen:2020}).

Further investigating these responses, in the following we solely discuss results for IllustrisTNG, but we note that SIMBA leads to similar conclusions. In principle, the responses of the \textit{1P} simulation set shown in Fig.~\ref{fig:pkee_z0_1P_IllustrisTNG_SIMBA} quantify the power spectrum dependence on CAMELS parameters. However, as described in Sec.~\ref{ssec:sims}, the CAMELS simulations cover a small cosmological volume and are therefore prone to being affected by cosmic variance, and we find this to be the case for the quantities considered in this work. 

Generally, we expect these fluctuations to have two contributions: a cosmological and an astrophysical contribution. The cosmological contribution quantifies realization differences due to initial conditions, while the second effect describes the associated stochasticity due to astrophysical processes on a wide range of scales: the strength of baryonic effects in a given realization will depend on its specific halo population, especially the abundance of massive halos for which feedback effects are expected to be strongest. For the remainder of this work, we will refer to the combination of these effects as cosmic variance, but it will be useful to keep this distinction in mind when interpreting our results.

\begin{figure*}
\begin{center}
\includegraphics[width=0.49\textwidth]{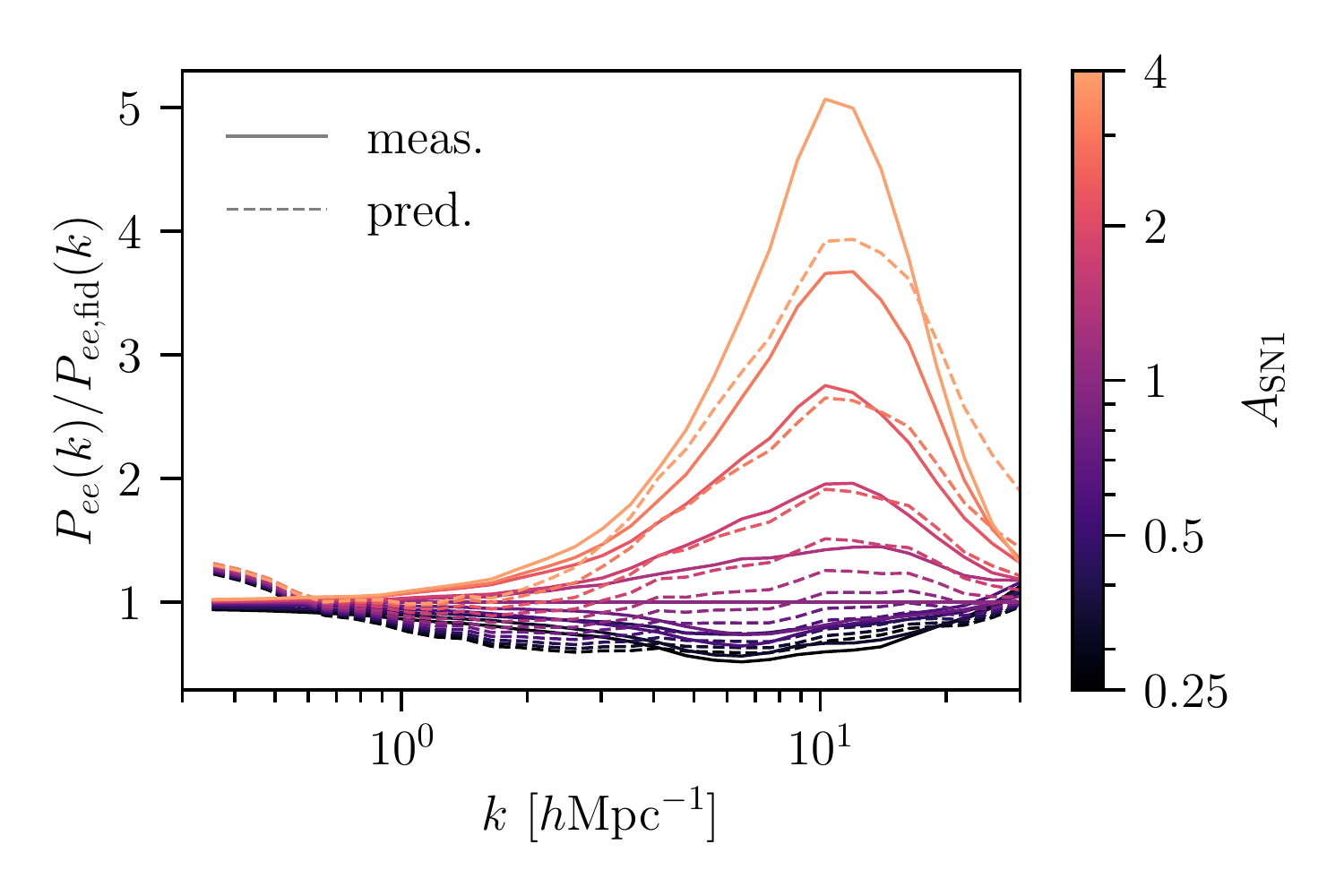}
\includegraphics[width=0.49\textwidth]{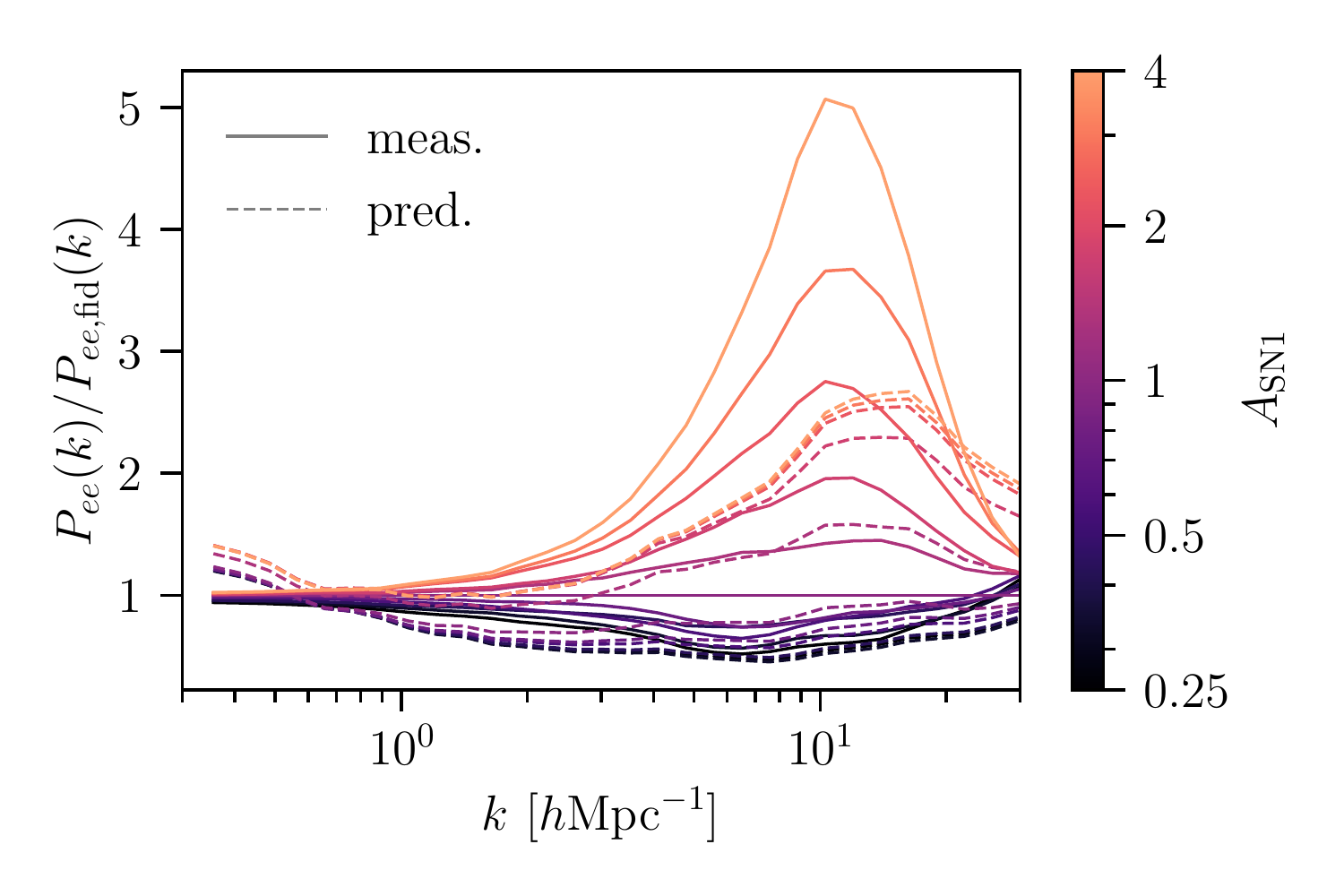}
 \caption{\textit{Left panel:} Comparison of the electron power spectra at redshift $z=0$ measured for the \textit{1P} set as a function of $A_{\mathrm{SN}1}$ and the emulated spectra from the Neural Network trained with the input parameter set $\boldsymbol{\vartheta} = (\Omega_{m}, \sigma_{8}, A_{\mathrm{SN}1}, A_{\mathrm{SN}2}, A_{\mathrm{AGN}1}, A_{\mathrm{AGN}2})$. \textit{Right panel:} Comparison of the electron power spectra at redshift $z=0.0$ measured for the \textit{1P} set as a function of $A_{\mathrm{SN}1}$ and the emulated spectra from the Neural Network trained with the input parameter set $\boldsymbol{\vartheta} = (\Omega_{m}, \sigma_{8}, \bar{f}_{\mathrm{bar}})$. The power spectra are shown only for IllustrisTNG and for different SNe feedback strengths; for clarity, we have normalized all of them by the fiducial measurement in CAMELS, which is defined by $A_{\mathrm{SN}1} =1$.}
\label{fig:pkme_z0_no-cv_As-vs-fbar}
\end{center}
\end{figure*}

The presence of cosmic variance fluctuations in the power spectra obtained from CAMELS might lead to inaccurate conclusions about the impact of baryonic feedback. To ensure that our results are dominated by physical effects rather than noise, we therefore first develop an emulator for obtaining noise-reduced power spectra from our data.
In this work, we choose to employ Neural Networks (NN) for this purpose, as Neural Networks effectively learn an approximation to the mean of an underlying function when trained to minimize the mean squared error loss (see e.g. Appendix C.3 in Ref.~\cite{Adler:2018}). Specifically, we use the \textit{LH} set of simulations to train a Neural Network to emulate the values of the power spectra $P_{ab}(k)$ as a function of cosmological and astrophysical parameters, i.e.
\begin{equation}
P_{ab}(k) = P_{ab}(k; \boldsymbol{\vartheta}),
\label{eq:pk_fit}
\end{equation}
where $a, b \in [e, m, g, p]$ and $\boldsymbol{\vartheta} = (\Omega_{m}, \allowbreak \sigma_{8}, \allowbreak A_{\mathrm{SN}1}, \allowbreak A_{\mathrm{SN}2}, \allowbreak A_{\mathrm{AGN}1}, \allowbreak A_{\mathrm{AGN}2})$. A detailed description of the employed methods can be found in Appendix \ref{ap:sec:pk_camels_meas}. The comparison between measurements and NN predictions for $P_{ee}(k)$ is shown in the left panel of Fig.~\ref{fig:pkme_z0_no-cv_As-vs-fbar} for the \textit{1P} set. For clarity, we have normalized all power spectra by the power spectrum of the fiducial model in CAMELS.
As can be seen, we find that while the NN fails to emulate the most extreme models of baryonic feedback in IllustrisTNG, it nevertheless provides a broad match to most measurements from the simulations. As we show in Appendix \ref{ap:sec:pk_camels_meas}, the observed differences are due mostly to cosmic variance, and we refer the reader to that section for a more detailed discussion of cosmic variance in numerical simulations.

Using the emulated, noise-averaged power spectra derived above, we can investigate the sensitivity of the electron density auto-correlation to baryonic feedback parameters in CAMELS. The left panel of Fig.~\ref{fig:pkme_z0_no-cv_As-vs-fbar} shows the dependence of the NN predictions on $A_{\mathrm{SN}1}$, varying within its prior range. As can be seen, we find the electron density power spectrum on intermediate and small spatial scales (large wave vectors) to increase strongly as a function of $A_{\mathrm{SN}1}$, whereas we see a sign for a reversal in this behavior on the smallest scales probed in our analysis. On larger spatial scales on the other hand, stellar feedback becomes ineffective and the cross-correlations are therefore mostly unaffected \cite{vanDaalen:2011}. We find that the remaining astrophysical parameters, i.e. $A_{\mathrm{SN}2}, A_{\mathrm{AGN}1}, A_{\mathrm{AGN}2}$, have a smaller impact on $P_{ee}(k)$, with responses mostly confined within cosmic variance uncertainties. The dominance of stellar feedback in determining baryonic effects in CAMELS has also been emphasized by Ref.~\cite{Villaescusa:2020a} and is likely due to the small volume covered by each simulation box. AGN feedback in particular is most effective in massive galaxies, which are severely underrepresented in small-volume simulations, thus leading to a possible underestimation of the true strength of AGN feedback in CAMELS. Additionally in Fig.~\ref{fig:pkme_z0_As-vs-fbar-vs-cv} in the Appendices, we show the dependence of $P_{ee}(k)$ on the two cosmological parameters varied in CAMELS, $\Omega_{m}$ and $\sigma_{8}$, highlighting that the electron power spectrum not only depends on astrophysics, but also on cosmology.

As noted above, we find the electron auto-correlation to increase with increasing SNe energy injection. This is somewhat counter-intuitive, as feedback effects are generally expected to decrease clustering amplitudes. However, our results agree with other works (\cite{Booth:2013, vanDaalen:2020}, Delgado et al., in prep.) that find a complex interplay between stellar and AGN feedback. As described in these analyses, it appears that lowering the efficiency of SNe feedback tends to boost the effects of AGN and thus increases the suppression of the matter power spectrum. We hypothesize this to hold even for CAMELS, despite the reduced strength of AGN feedback observed in the simulations: As discussed above, AGN feedback depends strongly on the availability of AGN-hosting halos in the simulations. We expect that lowering the strength of SNe feedback increases the probability for these halos to form, thus increasing AGN feedback strength. This mechanism might have stronger effects on the strength of AGN feedback than adapting the values of AGN-related parameters in CAMELS, possibly since it might lead to a higher probability for lower mass halos to host an AGN. Stronger stellar feedback on the other hand, might lead to a lower probability for forming massive halos that host an AGN, thus resulting in less suppression of the power spectra \cite{vanDaalen:2020}. In our case, this suggests that increasing SNe energy injection further reduces AGN feedback in CAMELS, thus leading to the increase in power spectrum amplitude with $A_{\mathrm{SN}1}$ observed in Fig.~\ref{fig:pkme_z0_no-cv_As-vs-fbar}.

At higher redshifts, $z=0.1, 0.47, 1.05$, the general trends in the dependence of $P_{ee}(k)$ on astrophysical parameters remain unchanged, but the overall impact of baryonic feedback increases as we move to later times\footnote{We do not show the results, as they are qualitatively similar to those obtained at $z=0$.}. These findings are consistent with the redshift-evolution of the matter power spectrum suppression due to baryons found in previous analyses (see e.g. Refs.~\cite{vanDaalen:2011, vanDaalen:2020}).

While we have only presented and discussed results for $P_{ee}(k)$, we have also investigated cross-correlations between the electron density and the matter and galaxy densities, as well as the electron pressure. We find that these statistics exhibit similar sensitivities to baryonic effects as $P_{ee}(k)$, albeit $P_{ge}(k)$ and $P_{pe}(k)$ being generally noisier. For $P_{ge}(k)$, this is probably due to the Poisson sampling of the galaxy density field, while we suspect that the higher noise in $P_{pe}(k)$ can be attributed to the pressure field being significantly affected by astrophysical stochasticity. For this reason, we do not present or discuss results for $P_{pe}(k)$ in the remainder of this work.

\subsection{Emulating power spectra from $\bar{f}_{\mathrm{bar}}$}\label{ssec:fbar_preds}

The astrophysical parameters $\boldsymbol{\vartheta}_{\mathrm{astro}} \coloneqq \{A_{\mathrm{SN}1}, A_{\mathrm{SN}2}, A_{\mathrm{AGN}1}, A_{\mathrm{AGN}2}\}$ varied in CAMELS control baryonic feedback strength by changing sub-grid modeling of the associated physical effects \cite{Villaescusa:2020a}. Therefore, these parameters are simulation-specific and it is difficult to translate their values between different simulations, in particular IllustrisTNG and SIMBA. This further implies that it is not directly possible to relate the CAMELS astrophysical parameters to observable quantities. In order to obtain constraints on physically-interpretable parameters, in this section we therefore aim to identify observable quantities that can act as surrogates for the CAMELS parameters. 

To this end, we use the results presented in Ref.~\cite{vanDaalen:2020} as a starting point: these findings suggest that the mean baryon fraction in massive halos is an accurate predictor for the matter power spectrum suppression due to feedback processes for different sets of simulations (see their Fig. 16). Investigating this further, we use CAMELS to test if the baryon fraction additionally allows us to predict the full shape of the power spectra considered in this analysis. Specifically, we collapse the feedback strength in each simulation into a single parameter, $\bar{f}_{\mathrm{bar}}$, which quantifies the mean baryon fraction in halos with masses $10^{12} \; h^{-1} \; M_{\odot} \leq M_{h} \leq 10^{13} \; h^{-1} \; M_{\odot}$ and is defined as
\begin{equation}
    \bar{f}_{\mathrm{bar}} = \frac{1}{N_{h}} \sum_{i}^
    {N_{h}}\frac{M_{\mathrm{bar}, h, i}}{M_{\mathrm{tot}, h, i}}~.
    \label{eq:fbar}
\end{equation}
In the above equation, $M_{\mathrm{bar}, h, i}$\footnote{Unless noted otherwise, we employ IllustrisTNG-native halo masses, derived using a friends-of-friends (FoF) algorithm \cite{Nelson:2019}.} denotes the total baryonic mass of halo $i$, $M_{\mathrm{tot}, h, i}$ its total mass, and $N_{h}$ is the number of halos with $10^{12} \; h^{-1} \; M_{\odot} \leq M_{h} \leq 10^{13} \; h^{-1} \; M_{\odot}$. We note that this definition of $\bar{f}_{\mathrm{bar}}$ is different from the one adopted in Ref.~\cite{vanDaalen:2020} as it constitutes an average over lower mass halos and excludes those with high masses. These halos are rare in CAMELS due to the small volume covered, and as we further describe in Sec.~\ref{ssec:constr_fdbck} they are additionally affected by stochasticity which leads to inaccurate values for $\bar{f}_{\mathrm{bar}}$ when including them in Eq.~\ref{eq:fbar}\footnote{We note that we additionally do not normalize $\bar{f}_{\mathrm{bar}}$ by the cosmic baryonic fraction, as this appears to lead to somewhat improved stability of our results.}.

With this definition of $\bar{f}_{\mathrm{bar}}$, we then proceed as outlined in Sec.~\ref{ssec:fdbck_effects} and Appendix \ref{ap:sec:pk_camels_meas} and train a NN to emulate $P_{ee}(k)$ using the reduced parameter set $\boldsymbol{\vartheta} = (\Omega_{m}, \sigma_{8}, \bar{f}_{\mathrm{bar}})$ as input. The obtained results are shown in the right panel of Fig.~\ref{fig:pkme_z0_no-cv_As-vs-fbar}. As can be seen, the NN predictions for this reduced parameter set are substantially worse than those for the full set of CAMELS parameters, particularly for models with high values of $A_{\mathrm{SN}1}$. Nevertheless, the emulated power spectra allow us to capture the basic trends in the responses. We can quantify the degradation in precision between the two models by comparing the mean relative deviation between measurements and predictions, defined as 
\begin{equation}
\Delta_{\mathrm{rel}} = \sqrt{\frac{1}{N_{k}N_{\mathrm{sim}}} \sum_{i, j}\frac{(P_{\mathrm{pred}, i}(k_{j})-P_{\mathrm{meas}, i}(k_{j}))^{2}}{P^{2}_{\mathrm{meas}, i}(k_{j})}},
\label{eq:Delta_Pk}
\end{equation}
where we average both over $N_{k}$ wave vectors $k_{j}$ with $1~h \mathrm{Mpc}^{-1} < k_{j} < 30~h \mathrm{Mpc}^{-1}$, and $N_{\mathrm{sim}}$ realizations $i$ for each data set used for testing\footnote{We exclude the smallest wave vectors (largest scales) from the relative deviation, as these scales are poorly fit by all models due to the significant cosmic variance, and would therefore dominate $\Delta_{\mathrm{rel}}$, thus biasing inferred differences.}. Evaluating $\Delta_{\mathrm{rel}}$ for the \textit{1P} and test sets, we find it to increase by roughly $30\%$ for the model based on $\bar{f}_{\mathrm{bar}}$, as compared to the one based on $\boldsymbol{\vartheta}_{\mathrm{astro}}$, thus further highlighting that the baryon fraction captures a significant, but not complete, amount of the information needed to predict baryonic feedback effects. A possible reason for the loss in predictive power when using $\bar{f}_{\mathrm{bar}}$ could be the near-constancy of this quantity for high values of $A_{\mathrm{SN}1}$, which is illustrated in Fig.~\ref{fig:pkee_z0_cc=fbar} in the Appendices.

We find that we can recover some of this lost information if we split $\bar{f}_{\mathrm{bar}}$ by a secondary halo property: specifically, we investigate the potential of $\bar{f}_{\mathrm{bar}}$ in three radial bins as a predictor for baryonic feedback. As we expect this quantity to be significantly affected by stochasticity due to the small volume covered by the CAMELS simulations, we additionally include a measure for cosmic variance (as described in Appendices \ref{ap:ssec:quant_stoch} and \ref{ap:ssec:account_stoch}) in our predictions. The results are shown in Fig.~\ref{fig:pkee_z0_cv_fbar-r} in the Appendices, and we find the mean relative deviation $\Delta_{\mathrm{rel}}$ to be comparable to the one based on $\boldsymbol{\vartheta}_{\mathrm{astro}}$. Furthermore, we additionally find slightly improved results when using the baryon fraction $\bar{f}_{\mathrm{bar}}$ for three mass bins, logarithmically-spaced between $10^{9} \; h^{-1} \; M_{\odot}$ and $10^{14} \; h^{-1} \; M_{\odot}$. These results suggest that some of the information lost can indeed be recovered when considering the baryon fraction as a function of halo mass or halo radius.

As the baryon fraction in halos can be constrained observationally (see e.g. Ref.~\cite{Gonzalez:2013}) and its value can be compared across different simulations, these results suggest that measurements of $\bar{f}_{\mathrm{bar}}$ could be used to model as well as constrain baryonic feedback strength in hydrodynamic simulations. 

\subsection{Constraints on cosmology and astrophysics} \label{ssec:constr_fdbck}

The results obtained in Sec.~\ref{ssec:fdbck_effects} show that the electron density power spectrum, $P_{ee}(k)$, at small spatial scales is sensitive to changes in baryonic feedback strength within a given hydrodynamic simulation\footnote{These conclusions also hold for the other power spectra considered, but we do not explicitly show these results.}. This opens up the interesting possibility of using these types of measurements to obtain simultaneous constraints on cosmology and astrophysics. In this section, we perform a simplified analysis and investigate the usefulness of such measurements by forecasting optimistic lower limits on attainable cosmological and astrophysical parameter constraints. We additionally derive an estimate for the systematic uncertainties associated to our incomplete knowledge of baryonic feedback effects.

\subsubsection{Statistical uncertainties}\label{ssec:stat_unc}

As can be seen from Fig.~\ref{fig:pkme_z0_no-cv_As-vs-fbar}, baryonic feedback mostly affects small-scale power spectra, but obtaining theoretical predictions and covariance matrices in this regime is challenging. In addition, CAMELS does not include enough simulations at the fiducial cosmology that would allow for estimating covariance matrices numerically. This therefore precludes a Fisher matrix analysis. We thus resort to a simplified analysis in order to forecast constraints on cosmological and astrophysical parameters from measurements of $P_{ee}(k)$. Specifically, we follow the methodology outlined in Ref.~\cite{Kendall:2017} and make the simplifying approximation that the joint posterior of the model parameters takes the form of an uncorrelated multivariate Gaussian. Under this assumption, Ref.~\cite{Kendall:2017} show that we can train a NN to infer the posterior means, $\mu_{i}$, and statistical (aleatoric) variances, $\sigma^{2}_{i}$, of the approximate posterior by setting the loss function of the NN to\footnote{We note that this approach is only valid if the NN constitutes an unbiased model of the data. We use the CV simulation set to verify that this is true for all cases considered in this analysis. Furthermore, to ensure positivity of the inferred posterior widths, we use the NN to constrain $\sigma_{i}^{2}$.}
\begin{equation}
    \mathcal{L}_{\mathrm{NN}} = \sum_{i} \left[ \frac{(\vartheta_{i}-\mu_{i})^{2}}{2 \sigma_{i}^{2}} + \frac{1}{2}\log{\sigma_{i}^{2}}\right],
\end{equation}
where $i$ runs over the number of model parameters\footnote{To test this approach, we repeat our analysis using the moment NN-based approach described in Ref.~\cite{Jeffrey:2020}: Specifically, we employ the loss function given in Ref.~\cite{Villaescusa:2020b}, finding very similar results. We therefore resort to our fiducial method for the remainder of this work.}. In this work, we employ power spectrum data as measured from CAMELS to train a NN using the above loss function, and infer the posterior means and variances of the cosmological and astrophysical parameters $\boldsymbol{\vartheta}$. In order to reduce dependence of our results on information contained in the smallest spatial scales observable with CAMELS, we use power spectrum data for $N_{k}=30$ wavevector values, logarithmically spaced between $k_{\mathrm{min}} = 0.36 \;h \; \mathrm{Mpc}^{-1}$ and $k_{\mathrm{max}} = 10\; h \; \mathrm{Mpc}^{-1}$. Following the discussion in Appendix \ref{ap:sec:pk_camels_meas}, we find it essential to include a measure for cosmic variance in a given simulation for obtaining accurate and precise parameter predictions from the NN. We refer the reader to this appendix for a more detailed discussion and a definition of the cosmic variance parameter. As before, we train the NN using data from the \textit{LH} set, subdividing it into the training, validation and test data sets as described in Appendix \ref{ap:sec:pk_camels_meas}. We note that we regard this simplified approach as a substitute for a Fisher matrix analysis to compute forecasted parameter uncertainties. Application to observational data will require simulation-based inference using well-motivated summary statistics, as well as methods such as neural density estimation \cite{Alsing:2019} to approximate the full parameter posterior.

\begin{figure*}
\begin{center}
\includegraphics[width=0.3\textwidth]{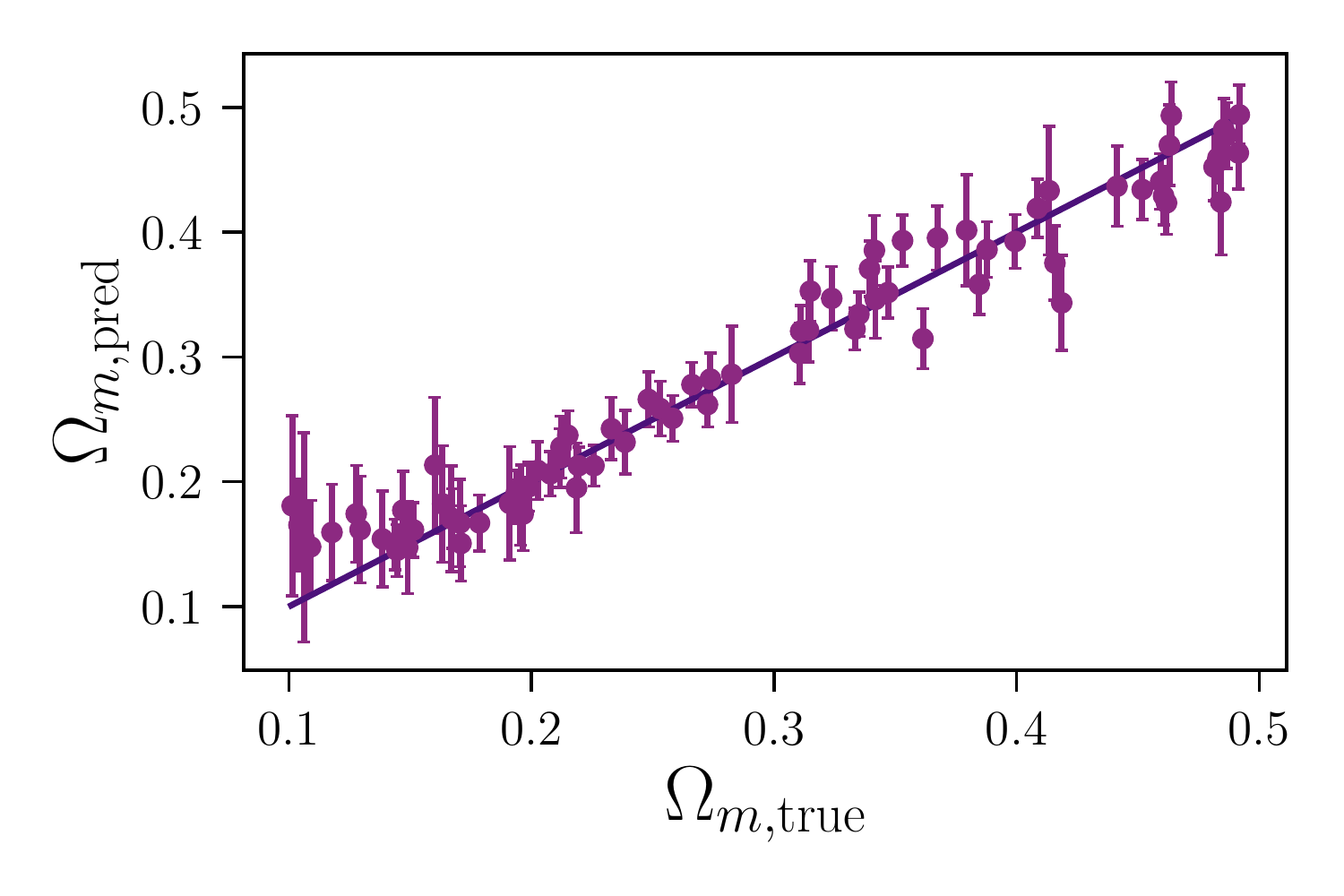}
\includegraphics[width=0.3\textwidth]{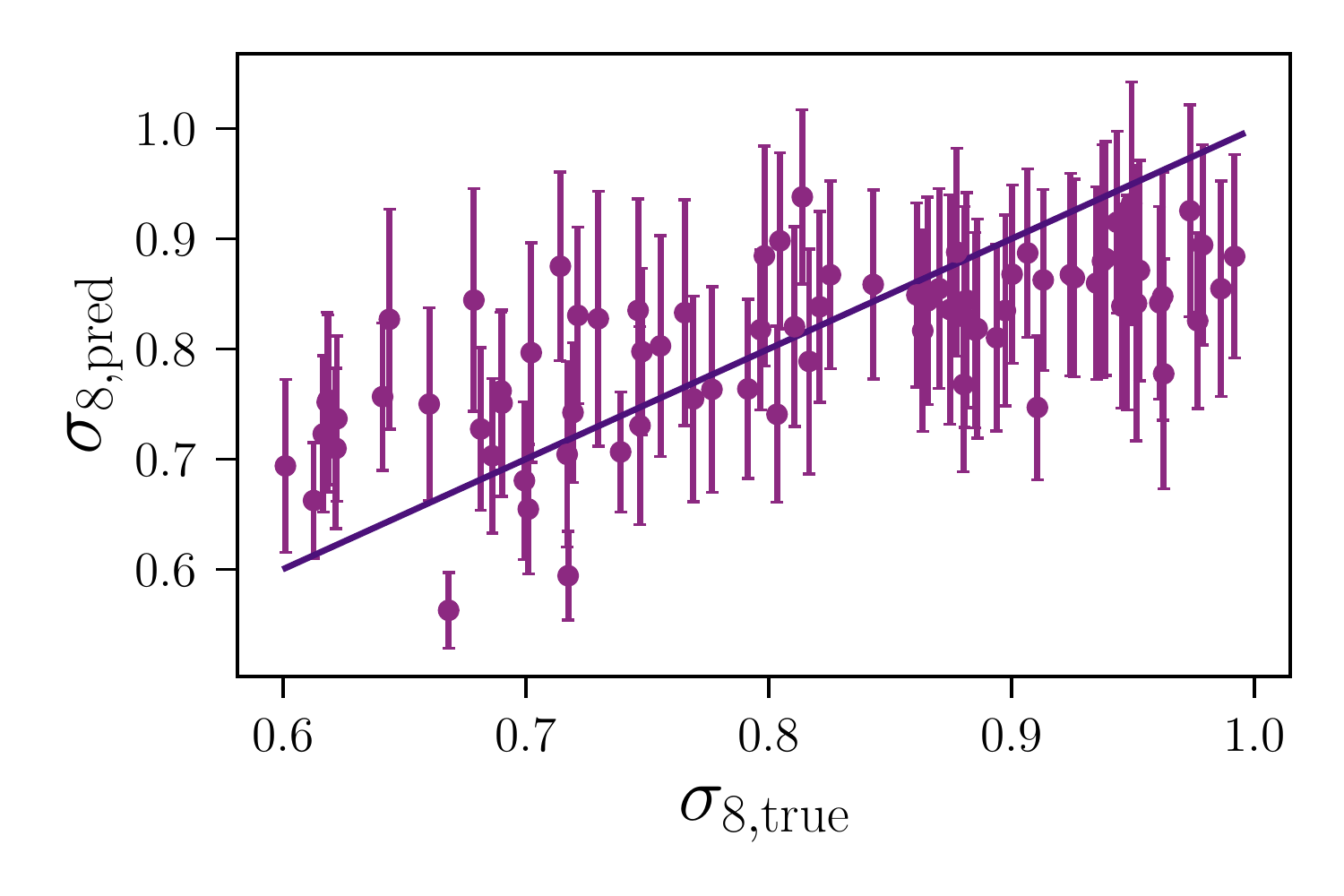}
\includegraphics[width=0.3\textwidth]{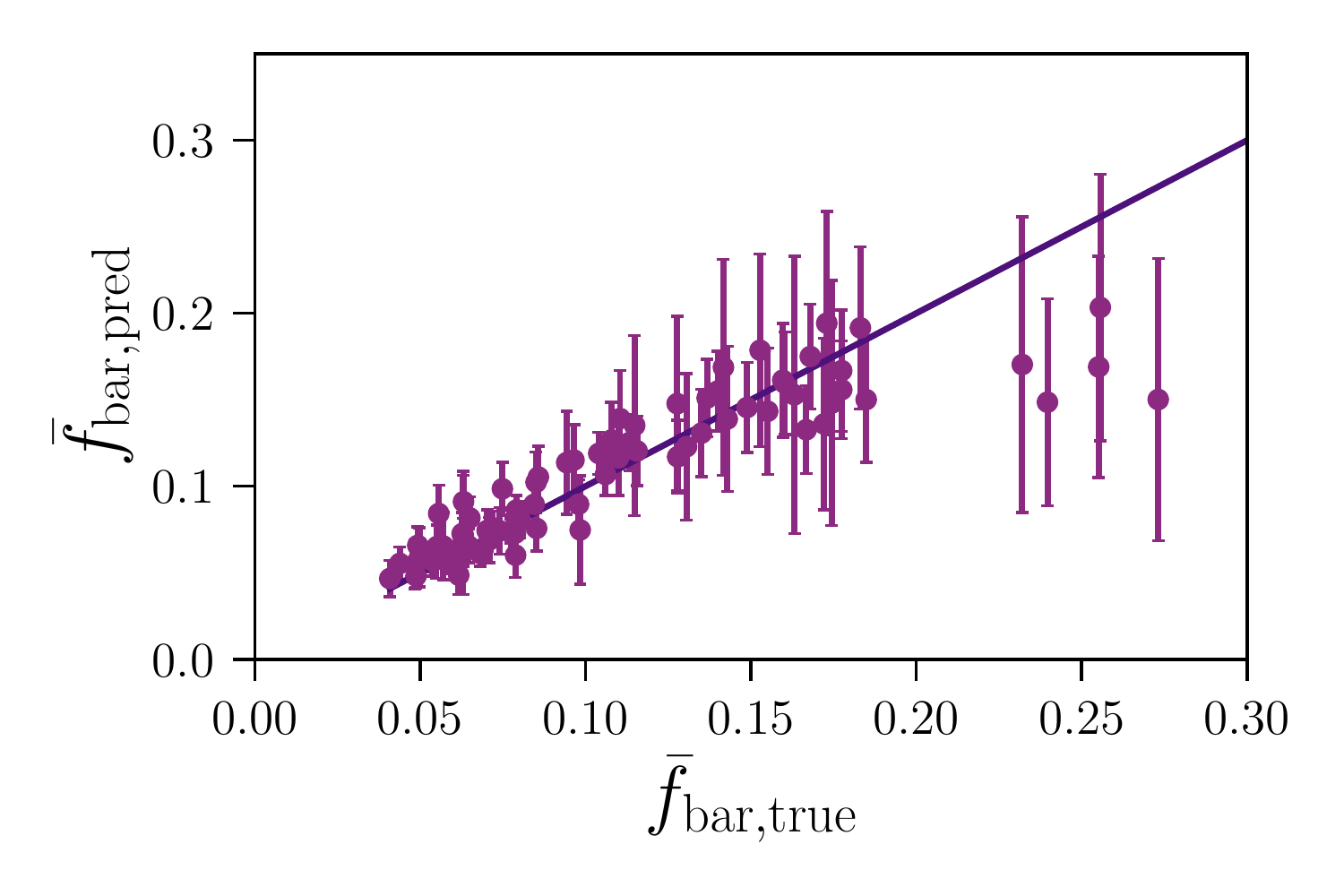}\\
\includegraphics[width=0.3\textwidth]{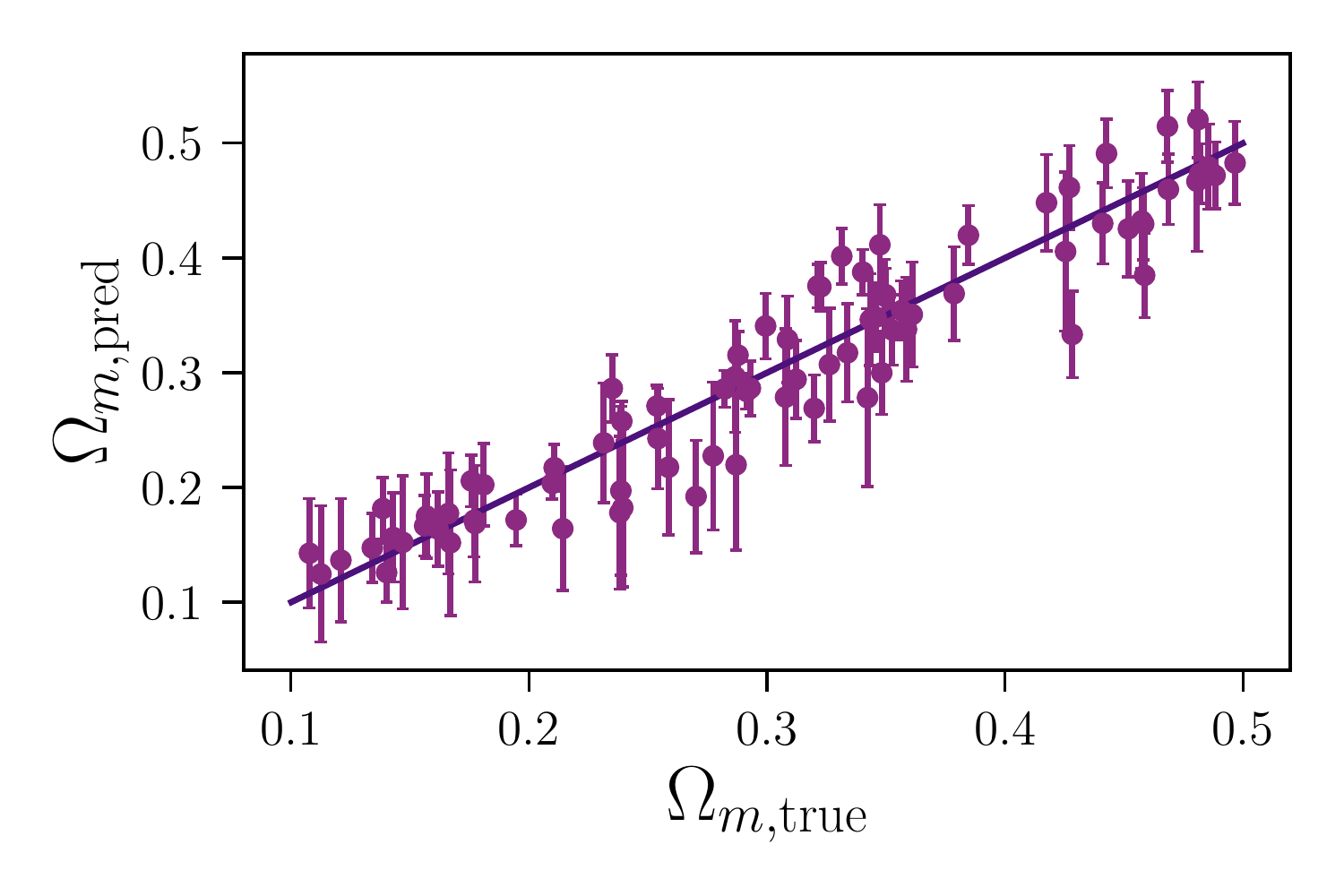}
\includegraphics[width=0.3\textwidth]{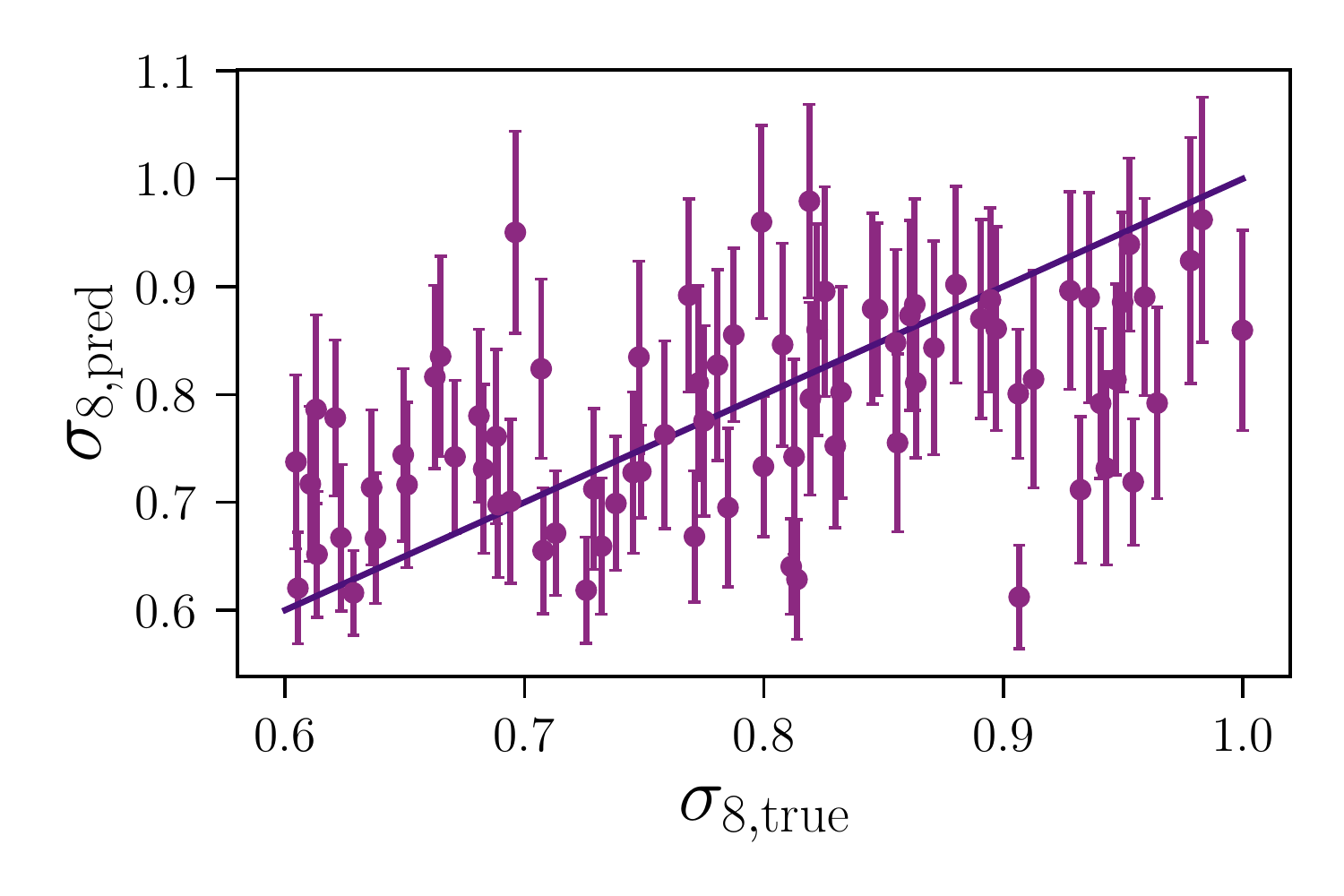}
\includegraphics[width=0.3\textwidth]{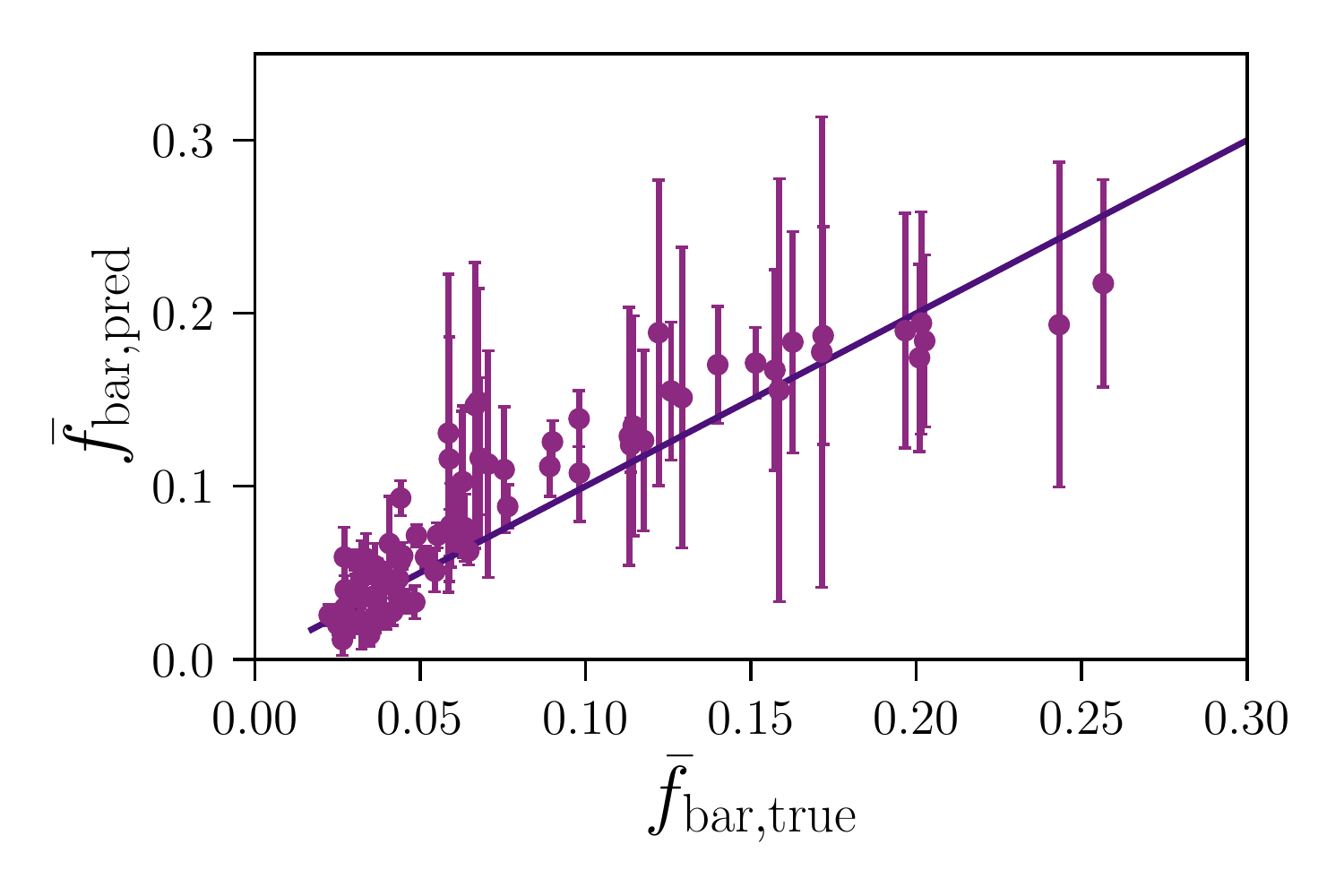}
 \caption{Comparison of NN-inferred values for CAMELS cosmological and astrophysical parameters obtained from $P_{ee}(k)$ at $z=0.0$ to their inputs, evaluated for a random draw of 80 simulations from the test set. \textit{Top panels:} The NN is trained and evaluated on the IllustrisTNG subset of CAMELS. This allows us to obtain an estimate for the statistical uncertainties associated with analyses of $P_{ee}(k)$. \textit{Bottom panels:} The NN is trained on IllustrisTNG, but evaluated on the SIMBA subset of CAMELS. This allows us to obtain an estimate for the systematic uncertainties in analyses of $P_{ee}(k)$, associated with our uncertain knowledge of baryonic physics.}
\label{fig:params_pk=mNe_z0_IllTNG-vs-SIMB}
\end{center}
\end{figure*}

As discussed in Sec.~\ref{ssec:fbar_preds}, baryonic feedback parameters in CAMELS are specific to a given simulation suite and cannot be directly linked to observations. It is therefore not possible to use the parameter set $\boldsymbol{\vartheta}_{\mathrm{astro}} = \{A_{\mathrm{SN}1}, A_{\mathrm{SN}2}, A_{\mathrm{AGN}1}, A_{\mathrm{AGN}2}\}$ to constrain baryonic physics in a simulation-independent way. However, the results shown in Sec.~\ref{ssec:fbar_preds} suggest that we can compress some of the information contained in these parameters into the observable quantity $\bar{f}_{\mathrm{bar}}$ given in Eq.~\ref{eq:fbar}, which is more readily transferable between different simulation suites. We therefore compute forecasted constraints on the reduced parameter set $\boldsymbol{\vartheta} = (\Omega_{m}, \sigma_{8}, \bar{f}_{\mathrm{bar}})$, instead of separately forecasting constraints on $A_{\mathrm{SN}1}, A_{\mathrm{SN}2}, A_{\mathrm{AGN}1}, A_{\mathrm{AGN}2}$.  

We obtain a measure for average parameter uncertainties by evaluating the trained NN on the \textit{CV} set; specifically, we compute the mean inferred variance, $\sigma_{i}^{2}$, for the \textit{CV} set and use it as our fiducial measure for predicted constraining power on parameter $i$. Using CAMELS $P_{ee}(k)$ power spectrum data in the wave vector range $[0.36, 10] \; h\mathrm{Mpc}^{-1}$ and at redshift $z=0$, we find that we can constrain $\Omega_{m}$ and $\bar{f}_{\mathrm{bar}}$, while $\sigma_{8}$ is largely unconstrained. This can be seen from the upper three panels of Fig.~\ref{fig:params_pk=mNe_z0_IllTNG-vs-SIMB}, which show the comparison between input and inferred model parameters for a set of 80 simulations, randomly drawn from the test set\footnote{As can be seen from Fig.~\ref{fig:params_pk=mNe_z0_IllTNG-vs-SIMB}, the NN predictions for high values of $\bar{f}_{\mathrm{bar}}$ appear biased. We attribute this to the limited number of \textit{LH} realizations with high baryon fractions, causing these to be underrepresented in our training set and thus causing biases in the NN inference. While the Latin Hypercube sampling ensures uniform sampling of the CAMELS parameter space, this is not the case for derived parameters such as $\bar{f}_{\mathrm{bar}}$.}. Specifically, we obtain forecasted relative uncertainties of roughly $7\%$ for $\Omega_{m}$ and $15\%$ for $\bar{f}_{\mathrm{bar}}$\footnote{We caution the reader that, while we find that the employed NN yields unbiased constraints when evaluated on the \textit{CV} set, it exhibits significant inaccuracies and biases for parameter values close to the prior bounds of the CAMELS suite. This is due to the fact that a NN trained to minimize the squared loss learns to predict the posterior mean (see e.g. Appendix A in Ref.~\cite{Villaescusa:2020b}), and might lead to overestimation of parameter uncertainties.}. As a comparison, we note that a similar analysis using the matter power spectrum, $P_{mm}(k)$, yields a relative error of roughly $4\%$ for $\Omega_{m}$, while we find $\bar{f}_{\mathrm{bar}}$ to be largely unconstrained. These forecasts use power spectra computed from the CAMELS simulation boxes, which cover a volume of $(25 \; h^{-1}\mathrm{Mpc})^{3}$. We therefore expect statistical uncertainties to be significantly reduced for larger cosmological volumes. However, this will not be true for systematic uncertainties, which we turn to next. 

We obtain similar results for the cross-power spectrum of the matter density and the electron number density field: we estimate relative uncertainties of roughly $6\%$ on $\Omega_{m}$, and $17\%$ on $\bar{f}_{\mathrm{bar}}$. The cross-correlation between the galaxy overdensity and the electron number density field on the other hand gives relative uncertainties of roughly double the size for $\Omega_{m}$, i.e. $11\%$, and $18\%$ on $\bar{f}_{\mathrm{bar}}$\footnote{We mainly attribute this loss in constraining power to the shot noise associated with observations of the galaxy distribution.}.

Comparing our results to the constraints reported in Ref.~\cite{Villaescusa:2021a}, we find the uncertainties reported in this work to be significantly larger. We mainly attribute this to the fact that the analysis in Ref.~\cite{Villaescusa:2021a} uses density fields as input, while we focus on power spectrum information only, and thus do not include any non-Gaussian information.

\subsubsection{Systematic uncertainties}

Inference from small-scale features of cosmic fields is significantly hampered by theoretical modeling uncertainties, and it is therefore crucial to test the robustness of the derived statistical parameter constraints by also estimating systematic uncertainties. At the scales considered in our analysis and for the hydrodynamic simulations employed, these uncertainties are mostly driven by our lack of knowledge of the effects of astrophysical processes and the related uncertainties on sub-grid models in numerical simulations. In order to obtain a rough estimate for the associated systematic uncertainties, we follow an empirical approach: as described in Sec.~\ref{sec:methods}, CAMELS consists of two sets of simulations, the IllustrisTNG and the SIMBA suites. So far, we have performed all analyses only for IllustrisTNG; SIMBA constitutes an independent set of simulations with a different implementation of baryonic feedback and a different approach for numerically solving the equations of hydrodynamics. As we have currently no reason to believe one of these simulations to provide a more accurate description of observational data than the other, we make the assumption that the differences in their predicted power spectra provide a rough measure for our systematic uncertainty on baryonic physics. This assumption gives us a way to test the robustness of the constraints derived above: if the derived constraints are robust to differences between the two simulations, it should be possible to use the NN model trained on IllustrisTNG to obtain unbiased parameter constraints when evaluating it on SIMBA. If on the other hand our results are sensitive to systematic differences between the two simulations, this will not be possible and the resulting biases can be taken as a measure for the systematic uncertainties associated with these analyses. We therefore assess the robustness of our results by evaluating the IllustrisTNG-trained NN model on SIMBA, and the results are illustrated in the lower three panels of Fig.~\ref{fig:params_pk=mNe_z0_IllTNG-vs-SIMB}. As before, these show the comparison between input and inferred model parameters for a set of 80 simulations, randomly drawn from the SIMBA \textit{LH} set. 

As can be seen, the recovered values of $\Omega_{m}$ and $\bar{f}_{\mathrm{bar}}$ appear to match their inputs reasonably well. We quantify potential biases, $b(\vartheta_{i})$, in recovering parameter $\vartheta_{i} \in (\Omega_{m}, \sigma_{8}, \bar{f}_{\mathrm{bar}})$ by computing the difference between the mean of the IllustrisTNG-trained model predictions for the SIMBA $CV$ set and the true value of the parameter, $\vartheta_{i, CV, \mathrm{fid}}$\footnote{For the \textit{CV} set, this is given by the CAMELS fiducial model.}, i.e.
\begin{equation}
    b(\vartheta_{i}) = \langle \hat{\vartheta}_{i} \rangle_{CV} - \vartheta_{i, CV, \mathrm{fid}},
\end{equation}
where $\langle \cdots \rangle_{CV}$ denotes the average over the \textit{CV} set. 
Evaluating this expression for the parameters that can be constrained using $P_{ee}(k)$ yields relative biases of $5\%$ for $\Omega_{m}$ and $8\%$ for $\bar{f}_{\mathrm{bar}}$. 
We estimate the statistical significance of these biases by comparing them to the uncertainties on the mean of the \textit{CV} set, i.e. $\sfrac{b(\vartheta_{i})}{\sigma(\bar{\vartheta_{i}})}$, where $\sigma(\bar{\vartheta_{i}})=\sfrac{\sigma(\vartheta_{i})}{\sqrt{N_{CV}}}$ and $N_{CV}$ denotes the number of \textit{CV} simulations. Evaluating this expression yields biases significant at levels of approximately $2 \sigma$ for both $\Omega_{m}$ and $\bar{f}_{\mathrm{bar}}$, i.e. they are not highly significant with respect to the error on the mean of the SIMBA \textit{CV} set. In addition, we can compare the biases to the corresponding parameter uncertainties given in Sec.~\ref{ssec:stat_unc}, and we find that the derived biases on $\Omega_{m}$ and $\bar{f}_{\mathrm{bar}}$ are slightly smaller than their predicted $1\sigma$ uncertainties. We are therefore led to conclude that analyzing the electron number density power spectrum, $P_{ee}(k)$, in the CAMELS volume yields constraints on $\Omega_{m}$ and $\bar{f}_{\mathrm{bar}}$ that appear rather robust to the differences in baryonic physics models implemented in IllustrisTNG and SIMBA. We note that this is not the case if we include high-mass halos in the computation of $\bar{f}_{\mathrm{bar}}$. Investigating this further, we find that the stellar-mass-to-halo-mass and the gas-mass-to-halo-mass relation in CAMELS both peak for halos with $10^{12} \; h^{-1} \; M_{\odot} \leq M_{h} \leq 10^{13} \; h^{-1} \; M_{\odot}$. This suggests that the halos used to compute $\bar{f}_{\mathrm{bar}}$ in Eq.~\ref{eq:fbar} may be less affected by baryonic feedback than their high-mass counterparts, thus making their baryonic fraction less susceptible to systematic differences in hydrodynamic simulations. 

Performing equivalent tests for $P_{me}(k)$ and $P_{ge}(k)$ on the other hand, we find a highly significant bias on the recovered values of $\bar{f}_{\mathrm{bar}}$. Recovering the fractional matter density $\Omega_{m}$ gives somewhat more acceptable results with relative biases of $6\%$ (significant at $3.2 \sigma$ with respect to the error on the mean) for $P_{me}(k)$, and $8\%$ (significant at $3.6 \sigma$ with respect to the error on the mean) for $P_{ge}(k)$.

In summary, these results suggest that analyses of $P_{ee}(k)$ on small spatial scales allow us to derive constraints on $\Omega_{m}$ and $\bar{f}_{\mathrm{bar}}$ that are largely robust to uncertainties due to baryon feedback as encoded in the differences between IllustrisTNG and SIMBA. Combining these findings with those derived in Sec.~\ref{ssec:fbar_preds}, we find that $\bar{f}_{\mathrm{bar}}$ is an acceptable predictor of baryonic feedback strength within a given hydrodynamic simulation (i.e. sub-grid physics model), furthermore generalizing to alternative feedback models. For analyses of $P_{me}(k)$ and $P_{ge}(k)$ on the other hand, we find that while the constraints on $\Omega_{m}$ are largely robust to baryonic physics uncertainties, the constraints on $\bar{f}_{\mathrm{bar}}$ are significantly affected by sub-grid modeling and thus dominated by systematic rather than statistical uncertainties. This might be due to the considered power spectra drawing information from different spatial scales, and thus exhibiting different sensitivities to sub-grid physics implementation. 

The works described in Refs. \cite{Villaescusa:2021a} and \cite{Villaescusa:2021b} performed a similar analysis using two-dimensional maps as input, finding the matter field to yield constraints on cosmological parameters robust to sub-grid modeling, while the constraints from all other fields do not generalize between IllustrisTNG and SIMBA. The comparison to our results therefore suggests that power spectra might be more robust to uncertainties due to baryonic feedback than two-dimensional field-level information. As before, we believe this to be related to these quantities being sensitive to information on different spatial scales.

\section{Caveats}\label{sec:caveats}

In the following, we discuss the caveats associated with our analysis, and outline possible avenues for future work.

First, the forecasts presented in this work assume measurements of $P_{ee}(k)$, a quantity that is not observable directly, but only through projected measurements such as the auto-correlation of the FRB dispersion measure field. Further work is required to translate the results obtained in this work to observables, while accounting for their inherent uncertainties due to observational noise, such as shot noise in FRB DM auto-power spectra \cite{Shirasaki:2017, Madhavacheril:2019, Takahashi:2021}. Furthermore, our analysis is based on power spectrum measurements at small spatial scales, which are affected by both theoretical and observational systematics. While we have quantified the susceptibility of our results to baryonic feedback uncertainties, several other systematics might hamper the usage of these scales for parameter inference. On the observational side, these include blending, intrinsic alignments or beam uncertainties (see e.g. Ref.~\cite{MacCrann:2017} for a discussion of small-scale systematics in WL measurements). In addition, all our results have been derived using a suite of hydrodynamic simulations that cover a limited volume and are thus potentially not fully converged. This is particularly relevant for the SIMBA simulations, in which AGN feedback has effects out to large spatial scales. 
Therefore, our results are possibly affected by inaccuracies due to finite box size, as well as being significantly affected by cosmic variance, as we have shown in Section \ref{sec:results} and Appendix \ref{ap:sec:pk_camels_meas}.

This analysis has highlighted some of the challenges faced when using hydrodynamic simulations of limited volume to investigate cosmological observables. While we have presented possible approaches to overcome these challenges, it will be insightful to perform similar analyses using for example a smaller suite of larger volume simulations. It will be interesting to compare these analyses to our results, and investigate if the methods presented in this work will be more readily applicable to these lower-noise simulations.
 
\section{Summary}\label{sec:conclusions}

The statistical power of current and future weak lensing surveys is limited by our uncertainties on the physics of baryons, and predictions for the effects of baryons on the matter distribution from hydrodynamic simulations currently differ significantly. In this work, we perform an idealized analysis and investigate the potential of the electron density power spectrum, $P_{ee}(k)$, to jointly constrain cosmology and baryonic feedback processes, which are mainly due to AGN and SNe. This quantity is not directly measurable but it underlies observations of FRBs and the kSZ. The constraints obtained from these analyses will be crucial for optimally benefiting from current and future weak lensing data, as they can be used to tune hydrodynamic simulations and thus improve theoretical models for the weak lensing signal. In order to investigate the dependence of $P_{ee}(k)$ on cosmology and astrophysics, we use the CAMELS suite of (magneto-)hydrodynamic simulations \cite{Villaescusa:2020a}. These simulations have been designed to serve as a data set allowing for quantification of our uncertain knowledge of baryonic physics as well as marginalization over these uncertainties. They therefore span a broad range in cosmological and astrophysical parameters, as well as two different sub-grid physics models, IllustrisTNG \cite{Nelson:2019} and SIMBA \cite{Dave:2019}. We find that $P_{ee}(k)$ on small spatial scales is sensitive to baryonic feedback effects, particularly to the SNe energy injection rate in IllustrisTNG, and galactic wind speed in SIMBA. Our results further show that the mean baryon fraction in intermediate-mass halos, $\bar{f}_{\mathrm{bar}}$, allows for reasonably accurate predictions of these dependencies and thus constitutes a promising observable quantity to constrain the effects of baryonic feedback in a simulation-independent way. 

We then forecast statistical uncertainties on cosmological and astrophysical parameters, focusing on the fractional matter density $\Omega_{m}$, the amplitude of matter fluctuations, $\sigma_{8}$, and $\bar{f}_{\mathrm{bar}}$. Our results suggest that analyzing $P_{ee}(k)$ at redshift $z=0$ allows us to constrain $\Omega_{m}$ and $\bar{f}_{\mathrm{bar}}$ to relative precision smaller than $15 \%$ for a CAMELS-like volume, while we find $\sigma_{8}$ to be unconstrained. We assess the robustness of these constraints to uncertainties in baryonic physics by deriving associated systematic uncertainties using an empirical approach based on the comparison of IllustrisTNG and SIMBA. This analysis suggests that $P_{ee}(k)$ is a promising observable for jointly constraining cosmology and astrophysics, largely robust to modeling differences in hydrodynamic simulations as implemented in IllustrisTNG and SIMBA. In contrast, we find the astrophysical constraints from the cross correlations of the electron density with the matter and galaxy density fields, $P_{me}(k)$ and $P_{ge}(k)$, to be sensitive to sub-grid modeling and thus to be dominated by systematic rather than statistical uncertainties.

Our analysis is subject to a number of caveats: First, it is based on three-dimensional power spectra, which are not directly observable, but only through line-of-sight projections of baryonic tracers of the LSS. In future work, it will be necessary to investigate how our results translate to these observables and their associated uncertainties due to observational noise. Furthermore, our analysis is based on power spectrum measurements at small spatial scales, which are prone to theoretical and observational systematics. While we have quantified the sensitivity of our results to baryonic feedback uncertainties, several other systematics might hamper the usage of these scales for parameter inference. Finally, our results have been derived using a suite of hydrodynamic simulations that cover a limited cosmological volume and are thus possibly not fully converged, which might affect our results. This work therefore presents an exploratory, proof-of-concept study, and it will be insightful to revisit our analysis once a similar suite of larger volume simulations becomes available. It will be particularly interesting to see if the performance of the methods presented in this work will be improved for simulations less affected by cosmic variance.

Despite these limitations, our results suggest that the electron power spectrum, $P_{ee}(k)$, provides a promising way to obtain rather robust constrains on cosmology and astrophysics. Furthermore, we find that the parameter $\bar{f}_{\mathrm{bar}}$ allows for emulating baryonic feedback strength within a given simulation, while also being largely robust to differences in sub-grid physics modeling, a result consistent with two related, recent analyses: Ref.~\cite{Giri:2021} presents an emulator for the matter power spectrum, based on the baryonicifcation model \cite{Schneider:2015, Schneider:2019}, and show that the power spectrum suppression due to baryons can be predicted using gas fraction measurements. The analysis in Ref.~\cite{Shirasaki:2021} presents a halo-model-based analysis of cross-correlations between FRBs and halos, finding these observables to be promising for jointly constraining cosmological and astrophysical parameters. Our results therefore provide further indication for the key role played by the baryon fraction in determining feedback strength \cite{vanDaalen:2020, Giri:2021, Shirasaki:2021}, even within the broad set of baryonic and sub-grid physics models considered in this work. It therefore appears promising to further pursue theoretical and observational studies of correlations between baryonic tracers of the LSS, alongside determining observational constraints on the baryon fraction. 

\acknowledgments

AN is grateful to the Institute for Advanced Study for the wonderful hospitality and the refuge during the time Princeton University was closed. AN would also like to thank the CCA for the hospitality and the opportunity to visit frequently. In addition, AN is happy to thank Simone Ferraro and Colin Hill for helpful discussions and feedback on an earlier version of this manuscript.
DAA was supported in part by NSF grants AST-2009687 and AST-2108944. AN and JD were supported through NSF grants AST-1814971 and AST-2108126. JD gratefully acknowledges support from the Institute for Advanced Study. The Flatiron Institute is supported by the Simons Foundation.

\appendix

\section{Computing cosmic fields from CAMELS} \label{ap:sec:comp_fields}

As described in Sec.~\ref{sec:methods}, we compute the galaxy overdensity field $\delta_{g}$, the matter overdensity field $\delta_{m}$, and the mass-weighted pressure field $p$\footnote{We note that we perform this additional mass-weighting to downweight low-mass voxels that feature unphysically large values for their pressure.}, in addition to $\delta_{e}$. We compute $\delta_{m}$ and $p$ directly from the simulation snapshots\footnote{We compute the electron pressure for each gas cell from its specific \texttt{InternalEnergy} $u$ using the ideal gas law $p = (\gamma -1)u \rho$.}. In order to compute $\delta_{g}$, we assume a survey loosely modeled on the Dark Energy Spectroscopic Instrument Bright Galaxy Sample (DESI BGS). We roughly follow Ref.~\cite{DESI:2016} and assume a constant comoving galaxy number density of $\bar{n} = 3 \times 10^{-3}~h^{3} \mathrm{Mpc}^{-3}$, which corresponds to an order of magnitude increase with respect to the SDSS LOWZ sample \cite{Parejko:2013}. Applying this number density constraint to the CAMELS simulations leaves us with an average of $N_{\mathrm{gal}} = 47$ galaxies in the CAMELS volume. We construct the overdensity field selecting the $N_{\mathrm{gal}}$ most massive galaxies in each simulation box in order to ensure a homogeneous sample. These low galaxy numbers are due to the small volume covered by a single CAMELS simulation box, and have motivated the creation of the CAMELS-SAM simulation suite (Perez et al., in prep.). CAMELS-SAM is a suite of 1000 N-body simulations that cover a volume of (100 $h^{-1}$ Mpc)$^3$ each, and have been run through a semianalytical model for galaxy formation, covering a broad cosmological and astrophysical parameter space. This suite therefore presents an addition to the original CAMELS data set, ideally suited to galaxy clustering measurements, and will be made publicly available (Perez et al., in prep.).

\section{Details on power spectrum measurements from CAMELS} \label{ap:sec:pk_camels_meas}

\subsection{Neural Network training and validation}

As described in Sec.~\ref{ssec:fdbck_effects}, we employ a NN to obtain emulated, noise-averaged power spectra for CAMELS. We use the LH set of simulations to train a Neural Network to emulate the values of the logarithm of the power spectra $P_{ee}(k)$, $P_{me}(k)$, $P_{ge}(k)$ and $P_{pe}(k)$ as a function of cosmological and astrophysical parameters. Specifically, we compute the power spectra on a discrete wavevector grid using $N_{k}=30$ wavevector values, logarithmically spaced between $k_{\mathrm{min}} = 0.36 \;h \; \mathrm{Mpc}^{-1}$\footnote{The minimal wave vector $k$ used in our analysis is determined by the volume covered by a single CAMELS simulation.} and $k_{\mathrm{max}} = 30\; h \; \mathrm{Mpc}^{-1}$, at redshifts $z=0, 0.1, 0.47, 1.05$ for all 1,000 simulations in the LH set. We split these data into training (700), validation (150) and test (150) set and train a simple NN for each power spectrum separately.  We build and train all NNs in this work using \href{https://pytorch.org/}{\tt{pytorch}}\footnote{\url{https://pytorch.org/}.}, and determine the optimal architecture using \href{https://optuna.org/}{\tt{optuna}}\footnote{\url{https://optuna.org/}.}. Our baseline architecture consists of a simple multilayer perceptron \cite{Hastie:2009}, i.e. 
\begin{enumerate}
    \item Input: Cosmological and astrophysical parameters, $N_{\mathrm{input}}$ values.
    \item Fully connected layer, $N_{\mathrm{neurons}}$ neurons.
    \item Leaky Rectified Linear Unit (LeakyReLU) activation function with fixed negative slope $ngslp = 0.2$.
    \item Dropout with rate $dr$.
    \item Output: Power spectrum array, $N_{\mathrm{output}} = 30$.
\end{enumerate}
Using \href{https://optuna.org/}{\tt{optuna}}, we adapt the number of layers (the number of times we stack steps 2, 3, and 4), the number of neurons per layer, $N_{\mathrm{neurons}}$, the dropout rate $dr$ per layer, the weight decay, and the learning rate to each specific case. The number of epochs used for training depends on the case considered, but we generally find a minimum of 1000 epochs to be sufficient, as the networks converge rather quickly.

In addition to evaluating the optimal model on the test set, we assess its performance on the 1P set of simulations. The comparison between model predictions and single realizations for the 1P power spectra differing only in the value of $A_{SN1}$ are shown in Fig.~\ref{fig:pkme_z0_no-cv_As-vs-fbar} for the electron auto-power spectrum. While we find broad overall agreement between measured and emulated power spectra, we do see residual differences, as shown in the left hand panel of Fig.~\ref{fig:pkme_z0_no-cv_As-vs-fbar}. Evaluating our model for the fiducial model in CAMELS and comparing to the CV realizations, we find the emulated power spectrum to match the mean of the CV set within cosmic variance, as can be seen from Fig.~\ref{fig:pkme_z0_sim=cv}. This suggests that the NN effectively averages over noise fluctuations in the CAMELS simulations, as intended, and emulates power spectrum means for a given set of model parameters. The discrepancies seen in Fig.~\ref{fig:pkme_z0_no-cv_As-vs-fbar} are therefore likely due to residual stochastic fluctuations in the single realizations\footnote{Alternatively, these residual differences could be due to the limited number of simulations used to train the NN. In order to test this hypothesis, we repeat our analysis for the parameter set $\boldsymbol{\vartheta} = (\Omega_{m}, \sigma_{8}, \bar{f}_{\mathrm{bar}})$ using a simpler model for the data: Specifically, we perform low-degree polynomial regression, and compare the best-fit models to the fiducial ones from the NN. We find the results for both cases to be very similar, and therefore conclude that the limitations of our training set are not the dominant reason for the observed differences between model predictions and data.}. 

\subsection{Quantifying cosmic variance in the simulations}\label{ap:ssec:quant_stoch}

\begin{figure}
\begin{center}
\includegraphics[width=0.45\textwidth]{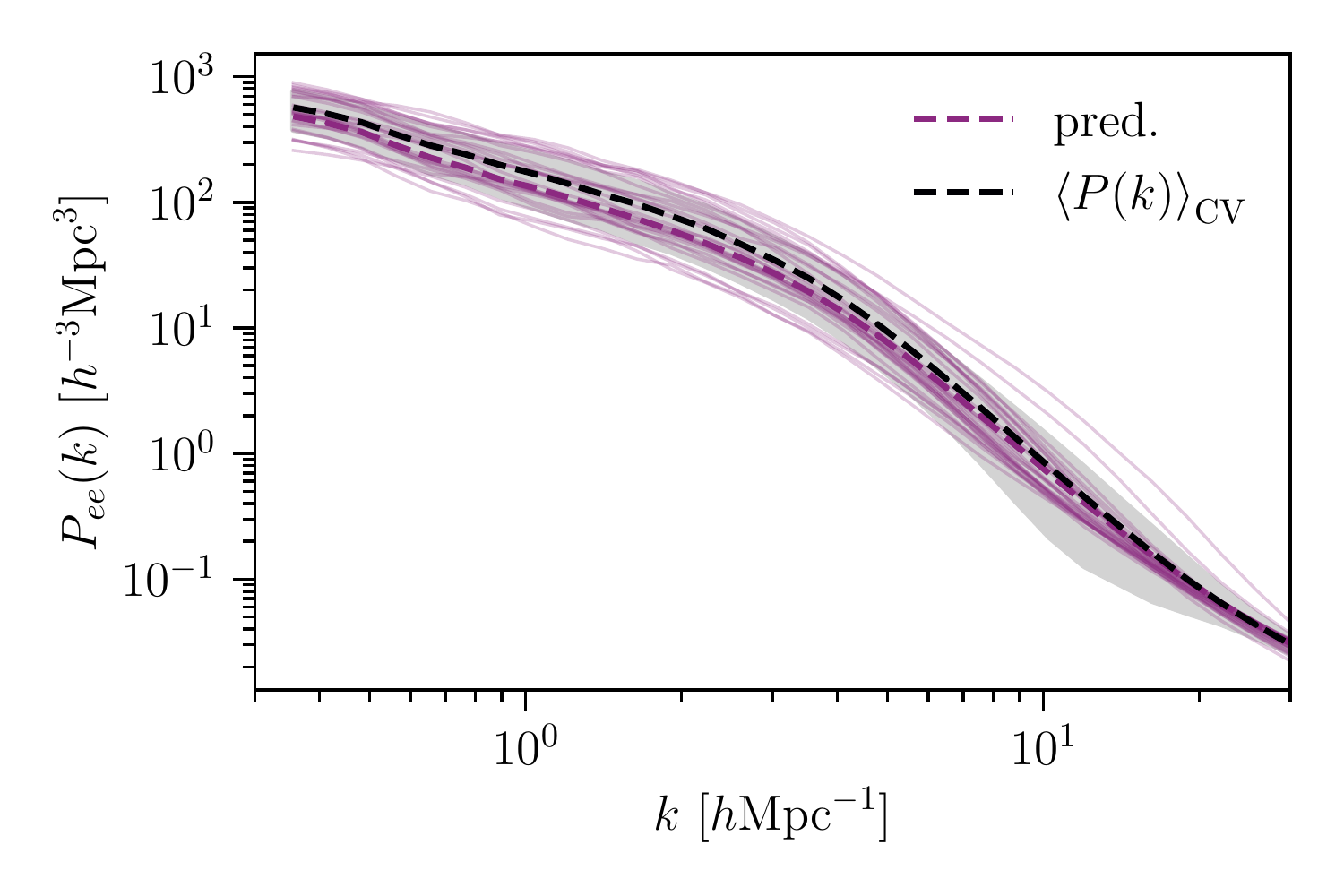}
 \caption{Comparison of the measured electron power spectra at redshift $z=0$, $P_{ee}(k)$, for the \textit{CV} simulations and the emulated spectrum for the CAMELS fiducial model derived using the parameter set $\boldsymbol{\vartheta} = (\Omega_{m}, \sigma_{8}, A_{\mathrm{SN}1}, A_{\mathrm{SN}2}, A_{\mathrm{AGN}1}, A_{\mathrm{AGN}2})$. The shaded region denotes the $1\sigma$ uncertainties derived from cosmic variance, and the transparent, solid lines show the single $CV$ realizations.}
\label{fig:pkme_z0_sim=cv}
\end{center}
\end{figure}

We can investigate this hypothesis and the potential source of these fluctuations using the $CV$ simulations. As described in Sec.~\ref{ssec:fdbck_effects}, we expect fluctuations in hydrodynamic simulations to be sourced by different initial conditions and the associated stochasticity in astrophysical processes. We further anticipate these latter fluctuations to be significantly correlated on the small spatial scales considered in our analysis, an effect that is borne out by the results shown in Fig.~\ref{fig:pkme_z0_sim=cv}. 

In order to further investigate these fluctuations, we focus on the electron power spectrum and consider its average value, $\bar{P}_{ee}$\footnote{Specifically, we compute the averaged power spectrum as $\bar{P}_{ee} \coloneqq \frac{1}{N_{k}} \sum_{i} P_{ee}(k_{i})$.}, in each \textit{CV} realization as a proxy thereof. We identify the physical quantities affecting $\bar{P}_{ee}$ by computing correlation coefficients for several quantities given in the simulations. Of all the quantities considered, we find the strongest correlation between $\bar{P}_{ee}$ and $\langle T_{\mathrm{gas}} \rangle_{M}$, which denotes the mass-averaged gas temperature in each simulation box\footnote{Specifically, we compute the gas temperature following the IllustrisTNG online documentation (\url{https://www.tng-project.org/data/docs/faq/}). We then obtain $\langle T_{\mathrm{gas}} \rangle_{M}$ by averaging the temperature over all gas voxels in a simulation box, weighted by voxel mass.}. We expect $\langle T_{\mathrm{gas}} \rangle_{M}$ to be affected by cosmic variance both through the particular halo mass function (HMF) realization as well as astrophysical stochasticity. The impact of these fluctuations on $\langle T_{\mathrm{gas}} \rangle_{M}$ and $\bar{P}_{ee}$ can be quantified in a back-of-the-envelope calculation: As the gas in the inter-halo medium is cold, we expect the mean gas temperature in a simulation box to be dominated by the temperature of gas residing in (massive) halos, which suggests that $\langle T_{\mathrm{gas}} \rangle_{M}$ should be proportional to the mean temperature of these halos. Assuming virial equilibrium and self-similar evolution, the temperature $T_{h}$ of a given halo is expected to scale with halo mass $M_{h}$ as (see e.g. Ref.~\cite{Borgani:2011})
\begin{equation}
\begin{aligned}
M_{h} \propto T_{h}^{\sfrac{3}{2}}.
\label{eq:M-T}
\end{aligned}
\end{equation}
From this relation, we obtain 
\begin{equation}
\langle T_{\mathrm{gas}} \rangle_{M} \propto M_{h}T_{h} \propto M_{h}^{\sfrac{5}{3}}. 
\label{eq:M-T}
\end{equation}
The reason for the additional factor of mass in the above equation is that the mass-average of the gas temperature $T_{\mathrm{gas}}$ is performed over the simulation box. As the total mass in each box and thus the normalization of the average are constant, the mass weight does not cancel, and we are left with the $M_{h}^{\sfrac{5}{3}}$-scaling derived above. Therefore, $\langle T_{\mathrm{gas}} \rangle_{M}$ is proportional to the mean integrated Compton-$y$ parameter of the halos in the simulation box.
As $\langle T_{\mathrm{gas}} \rangle_{M}$ is related to the average $M_{h}^{\sfrac{5}{3}}$ of that snapshot, for generality, we present all results in terms of this quantity. We denote it by $\mathcal{M}_{h, \mathrm{box}} \coloneqq \sfrac{1}{N_{h}} \sum_{i} M_{h, i}^{\sfrac{5}{3}}$\footnote{We note that $\mathcal{M}_{h, \mathrm{box}}$ does not have units of mass.}, but note that it physically provides a measure for the mean integrated Compton-$y$ parameter of clusters in a given simulation box. Combining with the results for $\bar{P}_{ee}$ discussed above, we finally obtain $\bar{P}_{ee} \propto \langle T_{\mathrm{gas}} \rangle_{M} \propto \mathcal{M}_{h, \mathrm{box}}$. The upper left panel of Fig.~\ref{fig:Tgas_M53} shows the correlation between $\mathcal{M}_{h, \mathrm{box}}$ and $\bar{P}_{ee}$ for the \textit{CV} set. The tightness of the correlation observed suggests that the fluctuations in $\langle T_{\mathrm{gas}} \rangle_{M}$, and thus $\bar{P}_{ee}$, are dominated by fluctuations in the halo mass function, and associated astrophysical stochasticity. We can see from Fig.~\ref{fig:Tgas_M53} that the correlation between $\mathcal{M}_{h, \mathrm{box}}$ and $\bar{P}_{ee}$ is not perfect, which we attribute to additional fluctuations not captured in our simple model. 

While these results have been explicitly derived only for the CAMELS simulations, we expect some of our findings to generalize to other simulations. In particular, the observed, tight relation between $\mathcal{M}_{h, \mathrm{box}}$ and $\bar{P}_{ee}$ allows us to predict the amplitude of cosmic variance fluctuations in a given simulation box using halo mass function information\footnote{We further test this hypothesis by training a NN to emulate the power spectra of the \textit{CV} set using only $\mathcal{M}_{h, \mathrm{box}}$ as input, finding good agreement between measurements and predictions and thus suggesting that $\mathcal{M}_{h, \mathrm{box}}$ indeed constitutes a reliable predictor for cosmic variance in the \textit{CV} set.}, which might be useful to understand cosmic variance effects in other simulations also. We note however that $\mathcal{M}_{h, \mathrm{box}}$ does not constitute an optimal predictor for cosmic variance, as we have not compared the performance of different measures in this work. 

\subsection{Accounting for cosmic variance in NN training}\label{ap:ssec:account_stoch}

These results suggest that individual power spectrum realizations are affected both by cosmological and astrophysical parameters as well as cosmic variance. Ideally, we could reduce the effects of cosmic variance on our results by learning the mapping between the electron and the Dark Matter distributions from CAMELS, and using it to inpaint baryons into larger-volume N-body simulations. In this work, we however take a simplified approach: from our results, we expect the performance of the NN to improve when additionally training on a measure for cosmic variance, such as $\mathcal{M}_{h, \mathrm{box}}$. In order to test this hypothesis, we train a NN to emulate $P_{ee}(k)$ given the cosmological and astrophysical parameters varied in CAMELS and $\mathcal{M}_{h, \mathrm{box}}$, and the results are shown in Fig.~\ref{fig:pkme_z0_cv}. These results constitute an improvement over the fits obtained not accounting for cosmic variance, as can be seen from comparing the left hand side of Fig.~\ref{fig:pkme_z0_no-cv_As-vs-fbar} with Fig.~\ref{fig:pkme_z0_cv}. We can quantify the improvement by comparing the mean relative deviation between measurements and model predictions, defined in Eq.~\ref{eq:Delta_Pk}; specifically we find an improvement of approximately $50 \%$.

\begin{figure*}
\begin{center}
\includegraphics[width=0.49\textwidth]{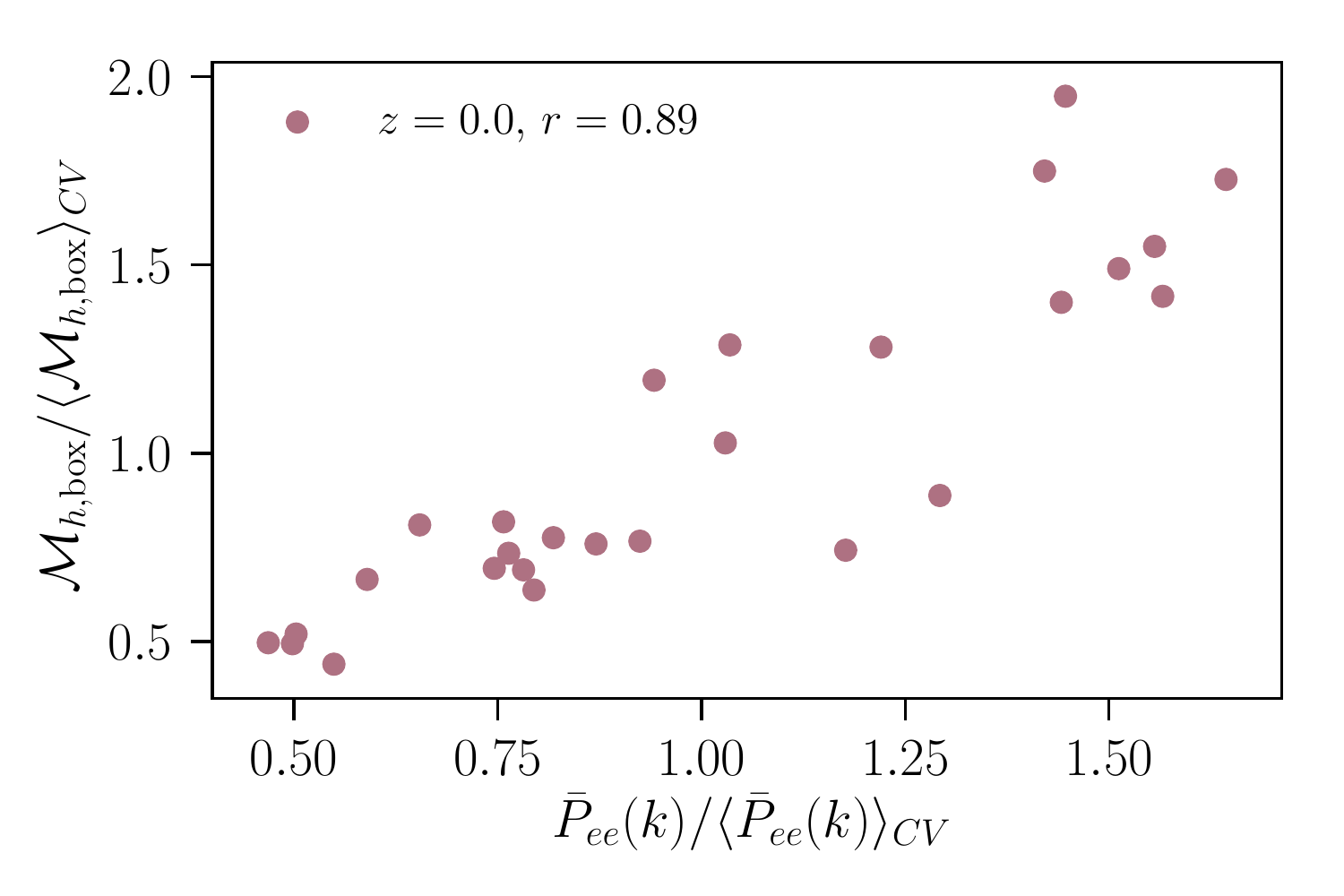}
\includegraphics[width=0.49\textwidth]{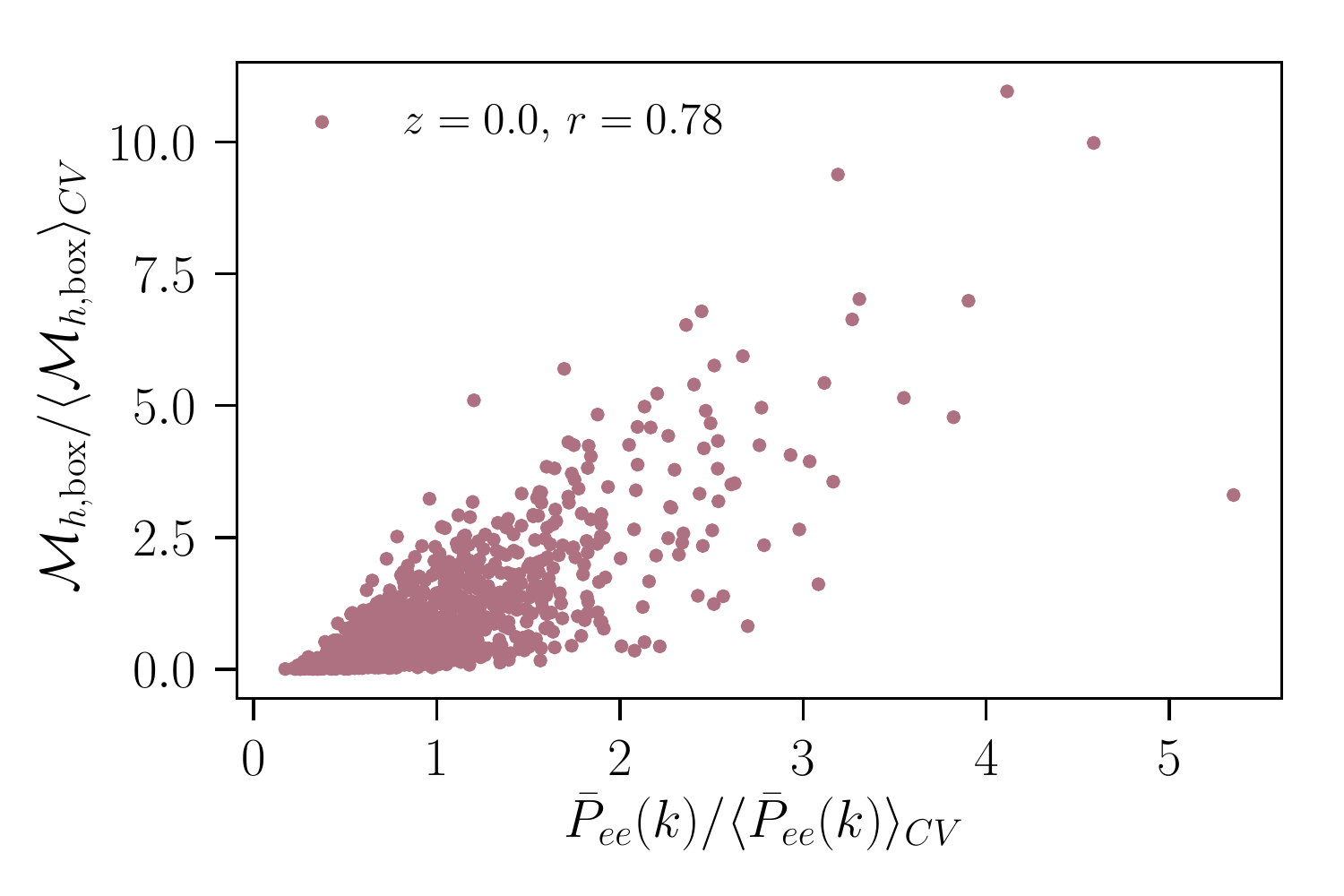}\\
\includegraphics[width=0.49\textwidth]{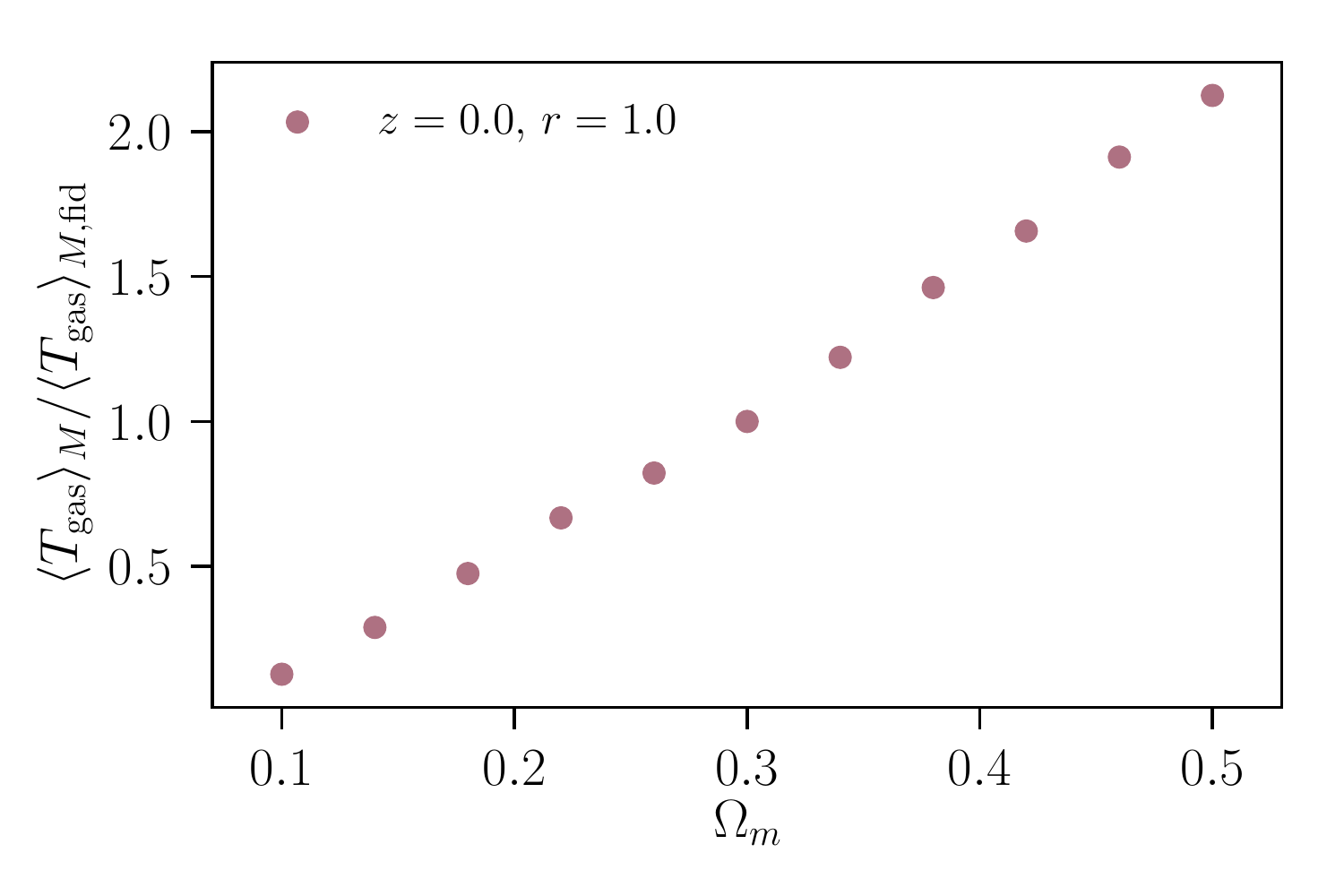}
\includegraphics[width=0.49\textwidth]{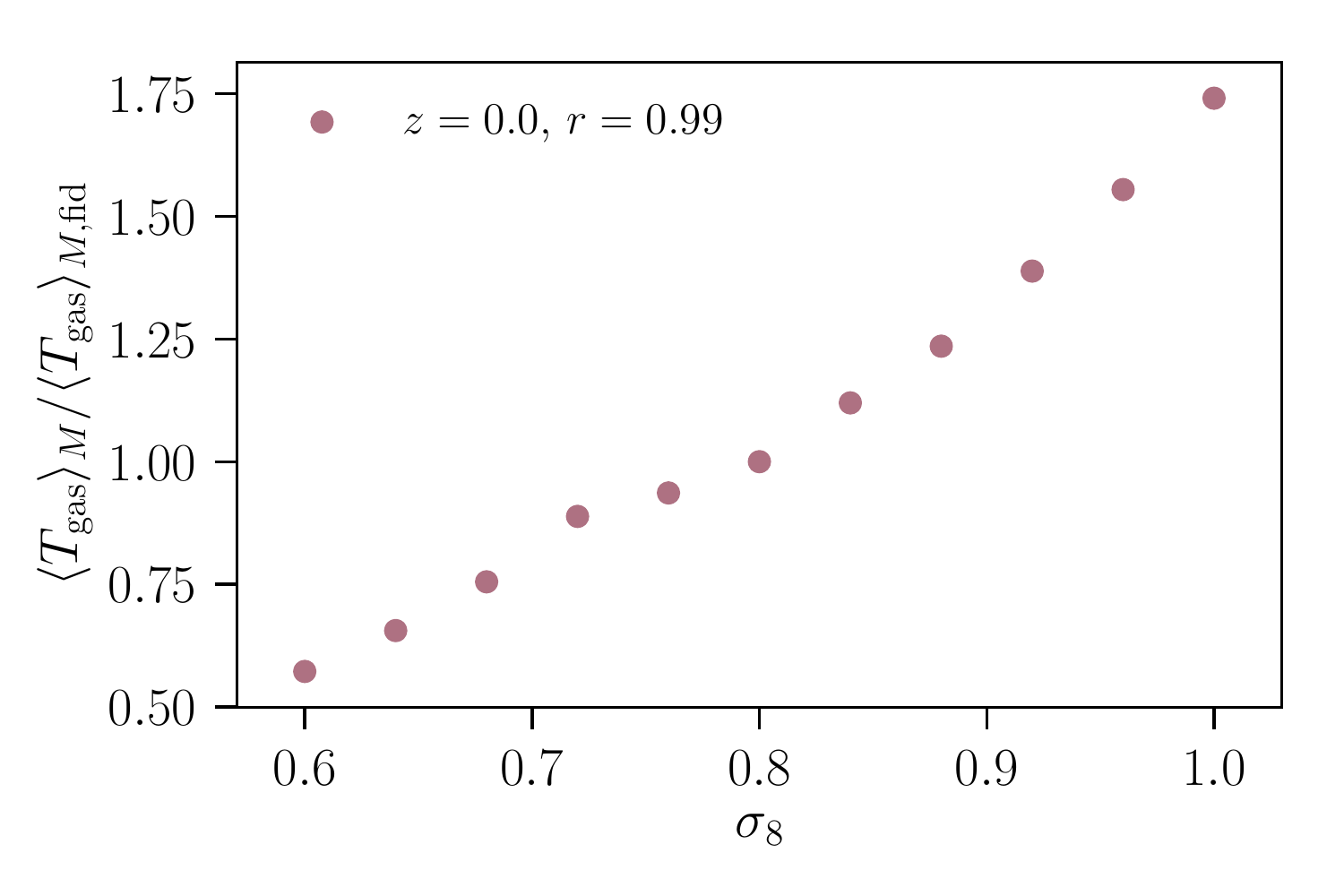}
 \caption{\textit{Upper panels:} Correlation between the cosmic variance parameter $\mathcal{M}_{h, \mathrm{box}}$ and $\bar{P}_{ee}$, for the \textit{CV} set on the left hand side, and for the \textit{LH} set on the right hand side. \textit{Lower panels:} Dependence of the average gas temperature, $\langle T_{\mathrm{gas}} \rangle_{M}$, on $\Omega_{m}$ and $\sigma_{8}$ for the \textit{1P} set.}
\label{fig:Tgas_M53}
\end{center}
\end{figure*}

\begin{figure}
\begin{center}
\includegraphics[width=0.49\textwidth]{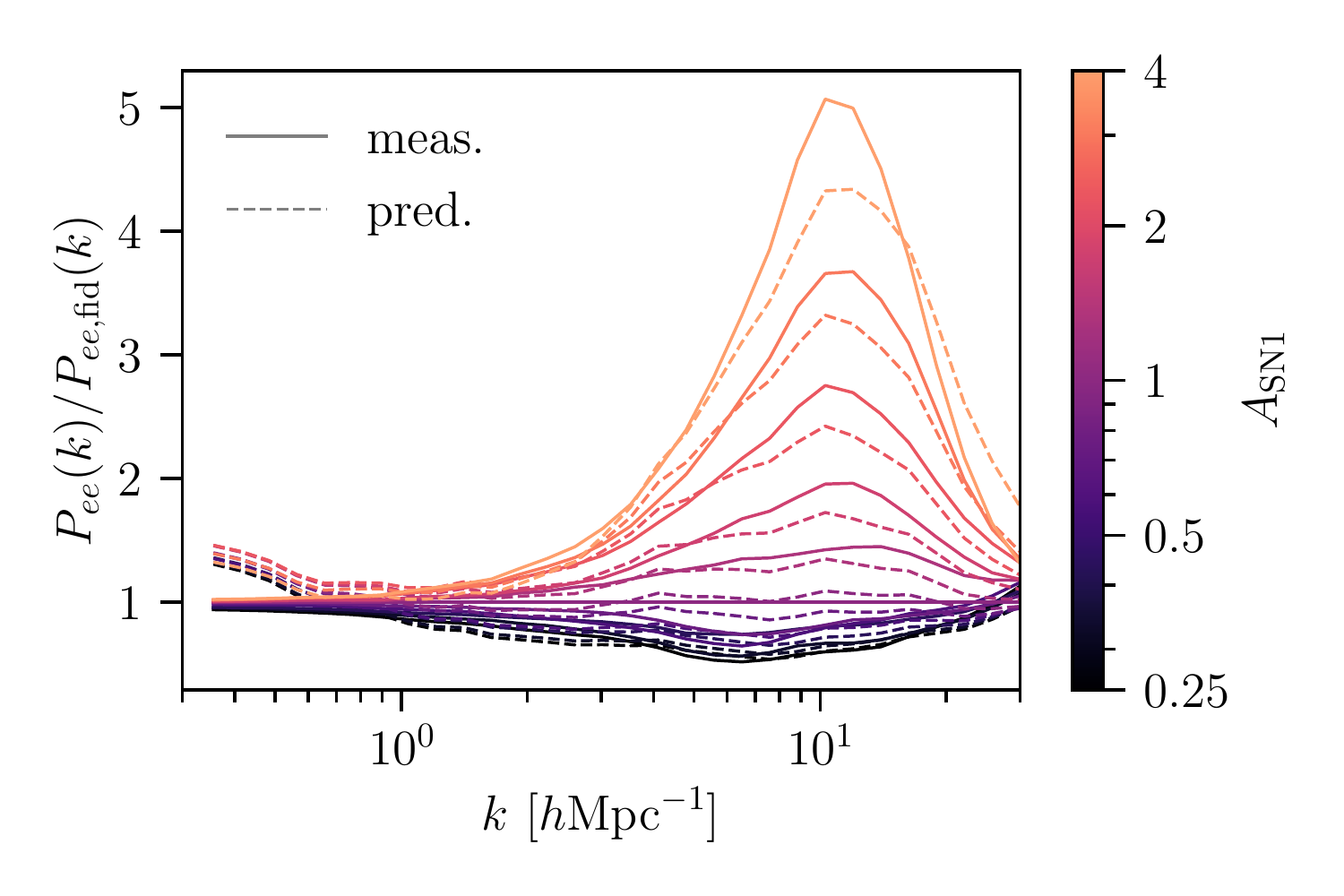}
 \caption{Comparison of the measured electron power spectra at redshift $z=0$, $P_{ee}(k)$, for the \textit{1P} simulations and the corresponding emulated spectra derived using the parameter set $\boldsymbol{\vartheta} = (\Omega_{m}, \sigma_{8}, A_{\mathrm{SN}1}, A_{\mathrm{SN}2}, A_{\mathrm{AGN}1}, A_{\mathrm{AGN}2})$ as well as the cosmic variance parameter $\mathcal{M}_{h, \mathrm{box}}$, as a function of SNe feedback strength $A_{\mathrm{SN}1}$. For clarity, we have normalized all power spectra by the fiducial measurement in CAMELS, which is defined by $A_{\mathrm{SN}1} =1$.}
\label{fig:pkme_z0_cv}
\end{center}
\end{figure}

These improvements are sizable but even after including $\mathcal{M}_{h, \mathrm{box}}$, there remain unexplained, residual differences between single realizations and emulated power spectra. Investigating this further, we consider the correlation between $\mathcal{M}_{h, \mathrm{box}}$ and $\bar{P}_{ee}$ for the \textit{LH} set and the results are shown in the upper right panel of Fig.~\ref{fig:Tgas_M53}. As can be seen, these quantities significantly decorrelate for the \textit{LH} set as compared to the \textit{CV} set. This behavior can be understood by looking at the lower panels in Fig.~\ref{fig:Tgas_M53}, which show the dependence of $\langle T_{\mathrm{gas}} \rangle_{M}$ on $\Omega_{m}$ and $\sigma_{8}$ as determined from the \textit{1P} set. We find that the average gas temperature additionally depends on cosmology and astrophysics\footnote{We discuss the physical origin of this dependence in Appendix \ref{ap:sec:Tgas_phys}.}, suggesting that it ceases to be a unique predictor of cosmic variance once CAMELS parameters and initial conditions are varied simultaneously. In these cases, $\mathcal{M}_{h, \mathrm{box}}/\langle T_{\mathrm{gas}} \rangle_{M}$ becomes a noisy predictor of cosmic variance, leading to the observed differences between the NN predictions and the measurements from the \textit{1P}/\textit{LH} sets of simulations. 

In summary, investigating the discrepancies between NN model predictions and the single realizations of $P_{ee}(k)$, we find that these differences indeed appear to be driven by correlated stochastic fluctuations in the observed power spectra and that the emulated power spectra from the NN provide a good fit to their noise-averaged means for a given set of input parameters $\boldsymbol{\vartheta} = (\Omega_{m}, \sigma_{8}, A_{\mathrm{SN}1}, A_{\mathrm{SN}2}, A_{\mathrm{AGN}1}, A_{\mathrm{AGN}2})$.

\section{Cosmological and astrophysical effects on average gas temperature} \label{ap:sec:Tgas_phys}

\begin{figure*}
\begin{center}
\includegraphics[width=0.49\textwidth]{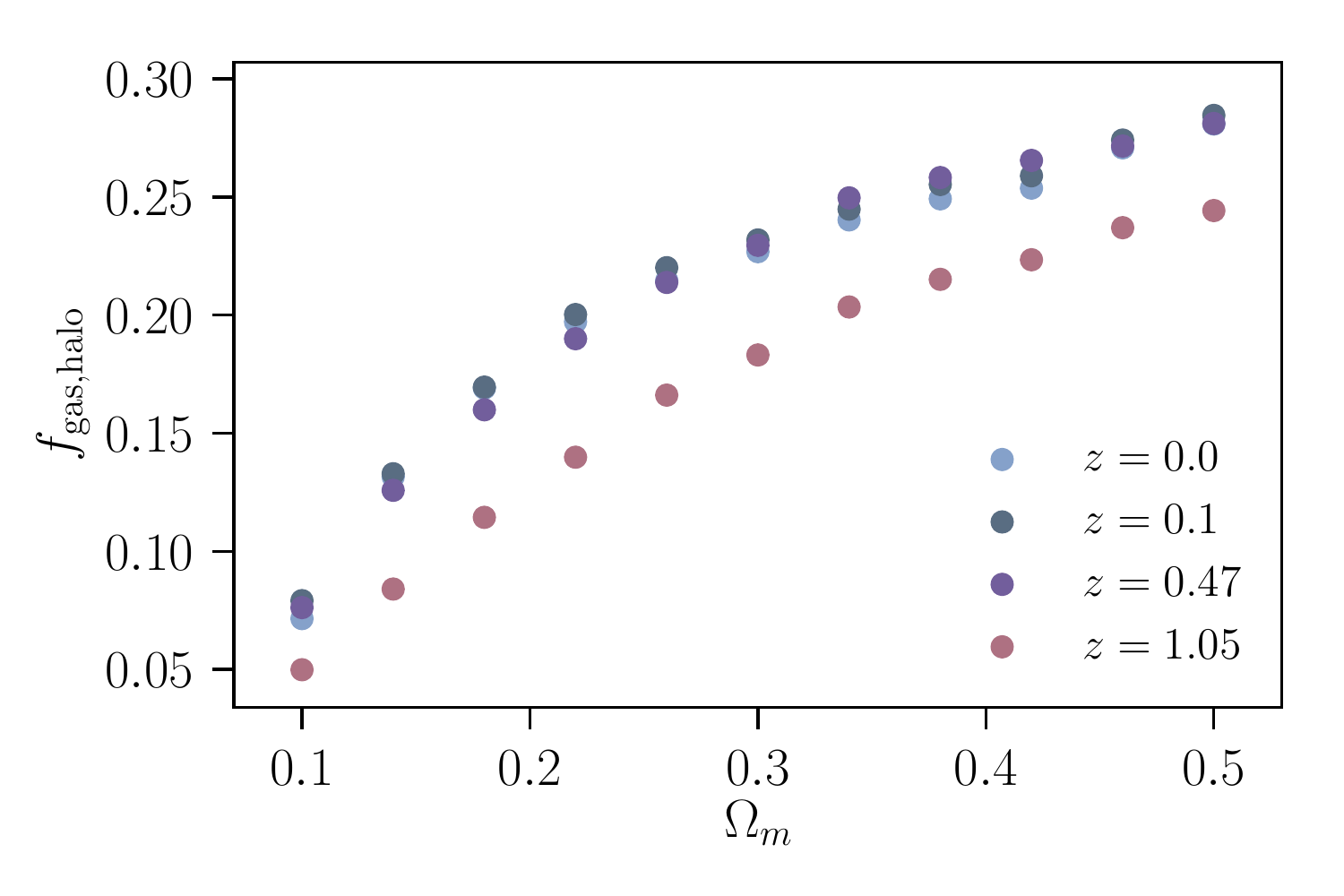}
\includegraphics[width=0.49\textwidth]{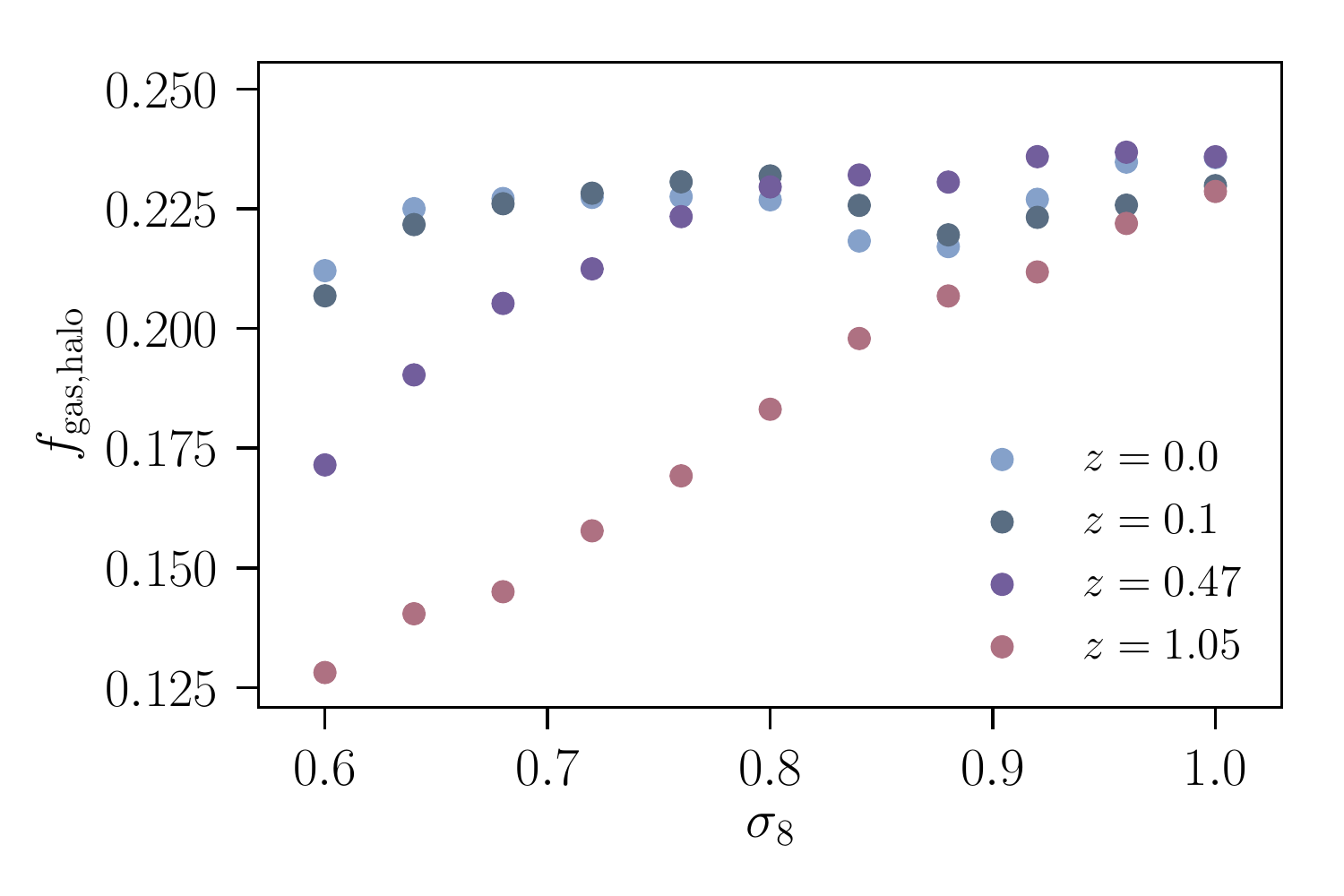}\\
\includegraphics[width=0.49\textwidth]{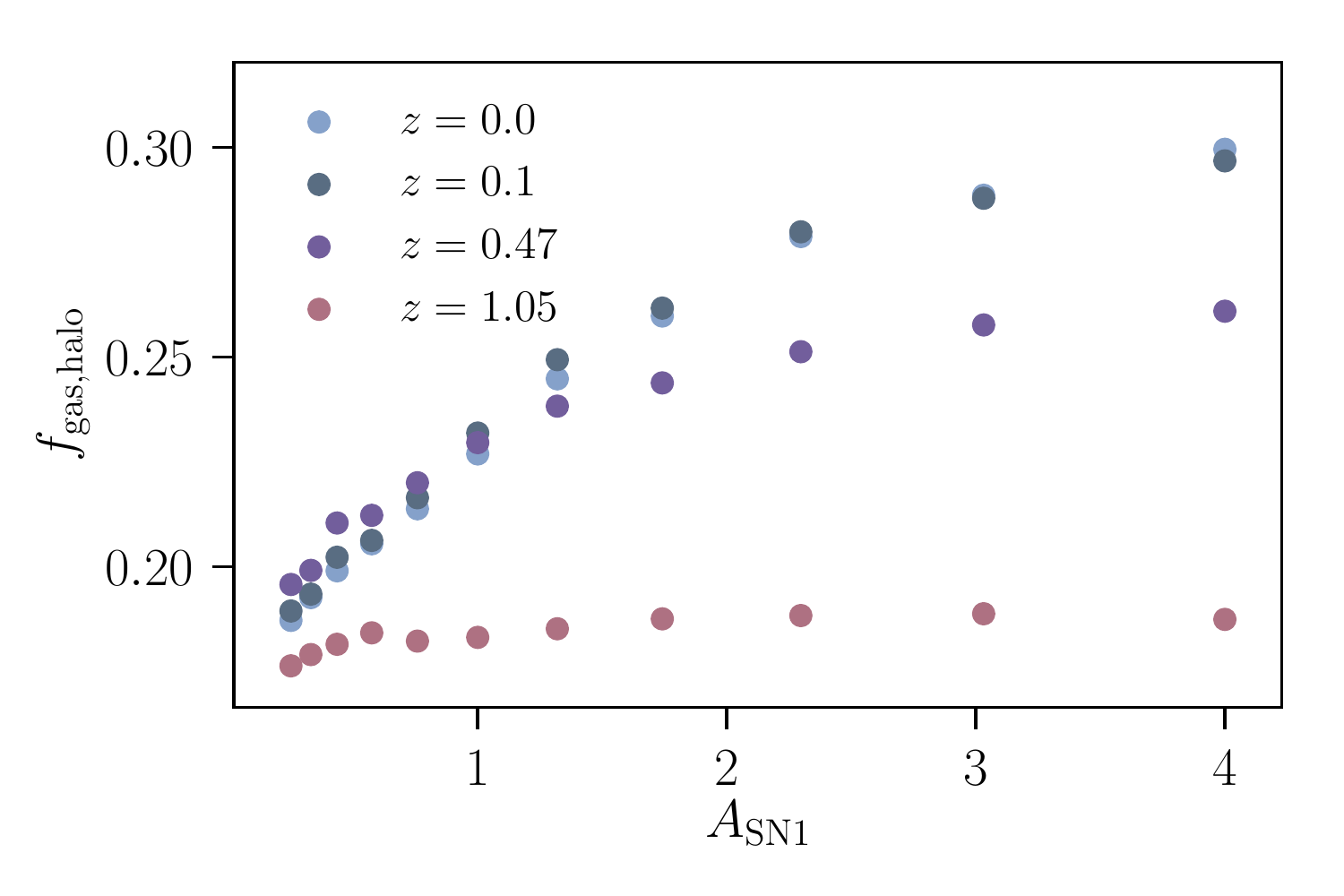}
\includegraphics[width=0.49\textwidth]{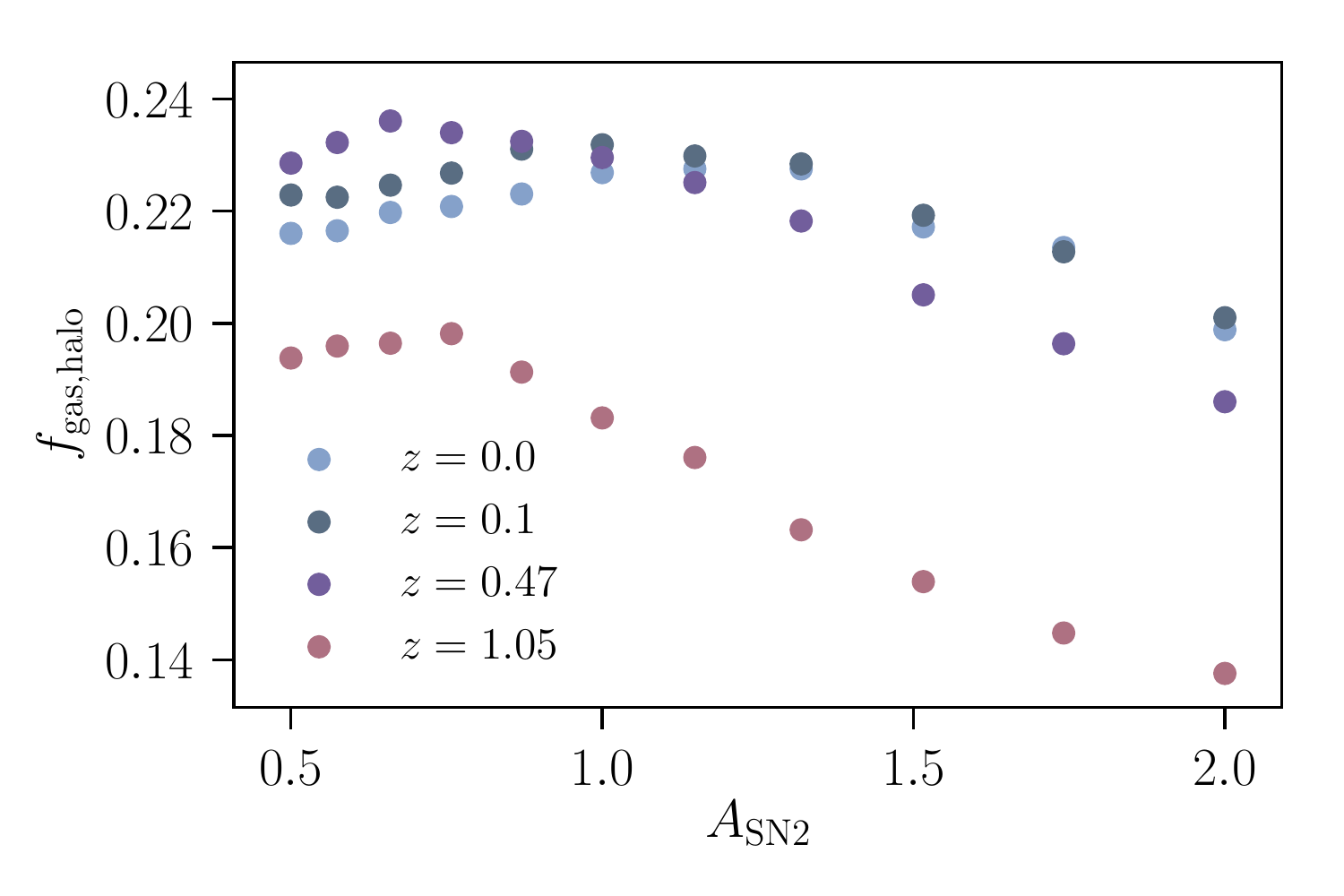}
 \caption{Dependence of the fraction of cosmic gas in halos, $f_{\mathrm{gas}, \mathrm{halo}}$, defined in Eq.~\ref{eq:f_gas} on cosmological and astrophysical parameters for the \textit{1P} set of simulations.}
\label{fig:fgas}
\end{center}
\end{figure*}

As described in Appendix \ref{ap:sec:pk_camels_meas} and illustrated in Fig.~\ref{fig:Tgas_M53}, we find the average gas temperature, $\langle T_{\mathrm{gas}} \rangle_{M}$, to strongly depend on cosmological parameters (and astrophysical parameters, not shown in Fig.~\ref{fig:Tgas_M53}). We use the \textit{1P} set of simulations to investigate the physical processes leading to these effects, as these simulations allow us to determine the response of a given quantity to separate changes in all CAMELS parameters. We expect the average gas temperature in a given simulation snapshot to be driven by the amount of gas inside halos, as the inter-halo gas is cold and will not significantly contribute to $\langle T_{\mathrm{gas}} \rangle_{M}$. Therefore, $\langle T_{\mathrm{gas}} \rangle_{M}$ should be sensitive to the number of halos in a simulation box and their gas content. An observable that encapsulates both of these effects is the fraction of gas in halos, which we define as
\begin{equation}
f_{\mathrm{gas}, \mathrm{halo}} \coloneqq \frac{\sum_{h} M_{\mathrm{gas}, h}}{M_{\mathrm{tot}, \mathrm{gas}}},
\label{eq:f_gas}
\end{equation}
where $M_{\mathrm{gas}, h}$ is the gas mass of a given halo $h$ and $M_{\mathrm{tot}, \mathrm{gas}}$ denotes the total gas mass in the snapshot\footnote{To compute $M_{\mathrm{gas}, h}$, we use spherical overdensity halos, as in Ref.~\cite{Villaescusa:2020a}. We note however, that we obtain similar trends when using FoF halos instead.}. Figure \ref{fig:fgas} illustrates the response of the gas fraction in halos to changes in the CAMELS parameters.
As can be seen, $f_{\mathrm{gas}, \mathrm{halo}}$ strongly depends on $\Omega_{m}, \sigma_{8}, A_{\mathrm{SN}1}$ and $A_{\mathrm{SN}2}$: higher values of $\Omega_{m}$ and $\sigma_{8}$ lead to increased clustering and thus deeper potential wells, increasing $f_{\mathrm{gas}, \mathrm{halo}}$ as more gas is bound inside the associated halos. Increasing the values of $A_{\mathrm{SN}1}$ and $A_{\mathrm{SN}2}$, suppresses gas ejection and star formation inside halos, leaving more of their baryonic content in the form of gas. Interestingly, for $A_{\mathrm{SN}2}$ we find that $f_{\mathrm{gas}, \mathrm{halo}}$ initially increases but then starts to decrease for high values of $A_{\mathrm{SN}2}$. As $A_{\mathrm{SN}2}$ controls the speed of stellar-feedback-driven galactic winds, we attribute this to the fact that at high parameter values, wind speeds become so effective as to expel gas from halos, thus leading to decreasing gas fractions.

The observed dependencies are very similar to those found for $\langle T_{\mathrm{gas}} \rangle_{M}$, as can be seen from comparing Figures \ref{fig:Tgas_M53} and \ref{fig:fgas}. This suggest that the main reason for the dependence of $\langle T_{\mathrm{gas}} \rangle_{M}$ on cosmological and astrophysical parameters in CAMELS is that these affect the baryon fraction in halos, which in turn influences the average gas temperature.

\begin{figure*}
\begin{center}
\includegraphics[width=0.49\textwidth]{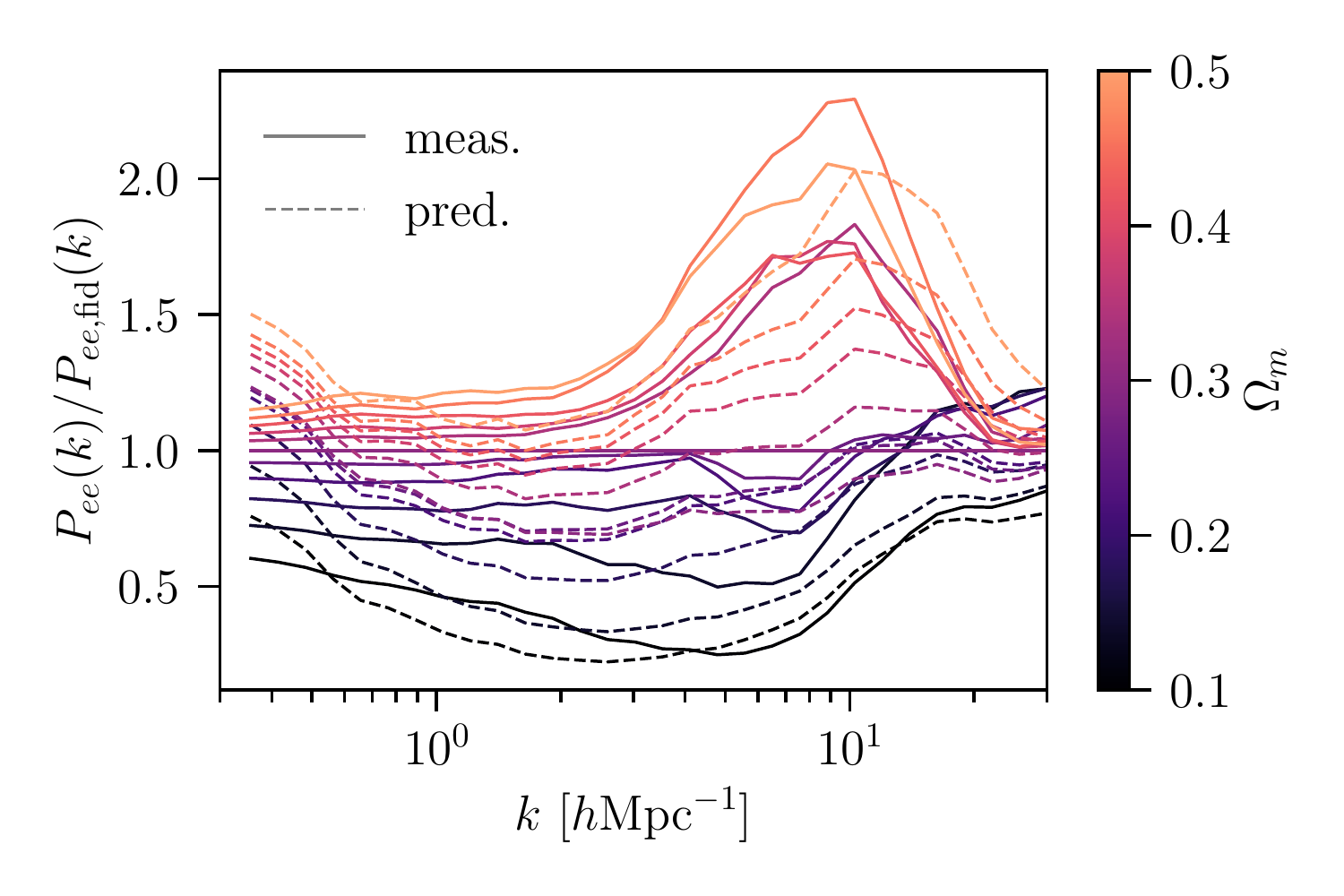}
\includegraphics[width=0.49\textwidth]{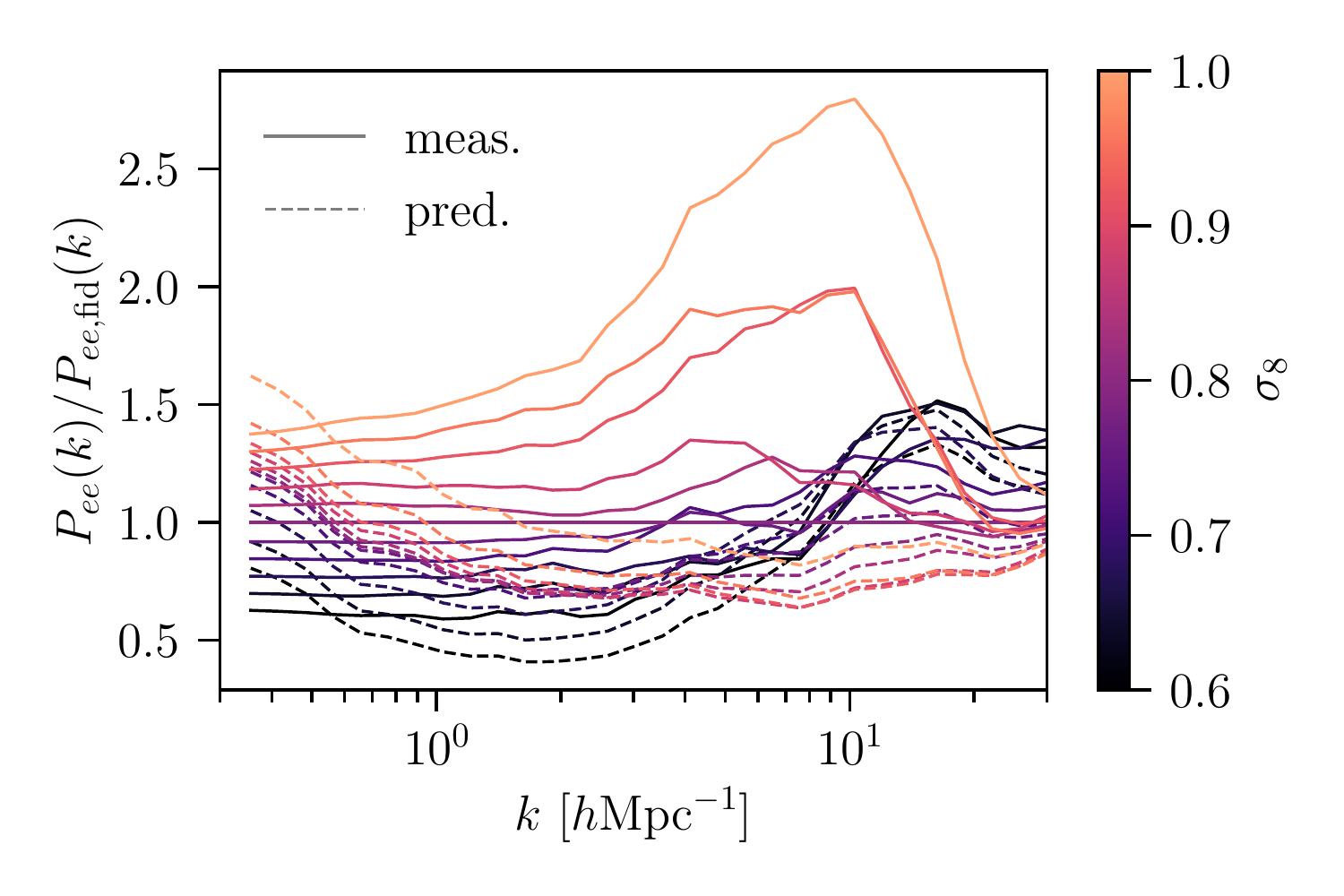}
 \caption{Comparison of the measured electron power spectra at redshift $z=0$, $P_{ee}(k)$, for the \textit{1P} simulations and the corresponding emulated spectra derived using the parameter set $\boldsymbol{\vartheta} = (\Omega_{m}, \sigma_{8}, \bar{f}_{\mathrm{bar}})$. The power spectra are shown for different values of $\Omega_{m}$ and $\sigma_{8}$, and for clarity, we have normalized all of them by the fiducial measurement in CAMELS, which is defined by $\Omega_{m}=0.3$ and $\sigma_{8}=0.8$.}
\label{fig:pkme_z0_As-vs-fbar-vs-cv}
\end{center}
\end{figure*}

\begin{figure}
\begin{center}
\includegraphics[width=0.49\textwidth]{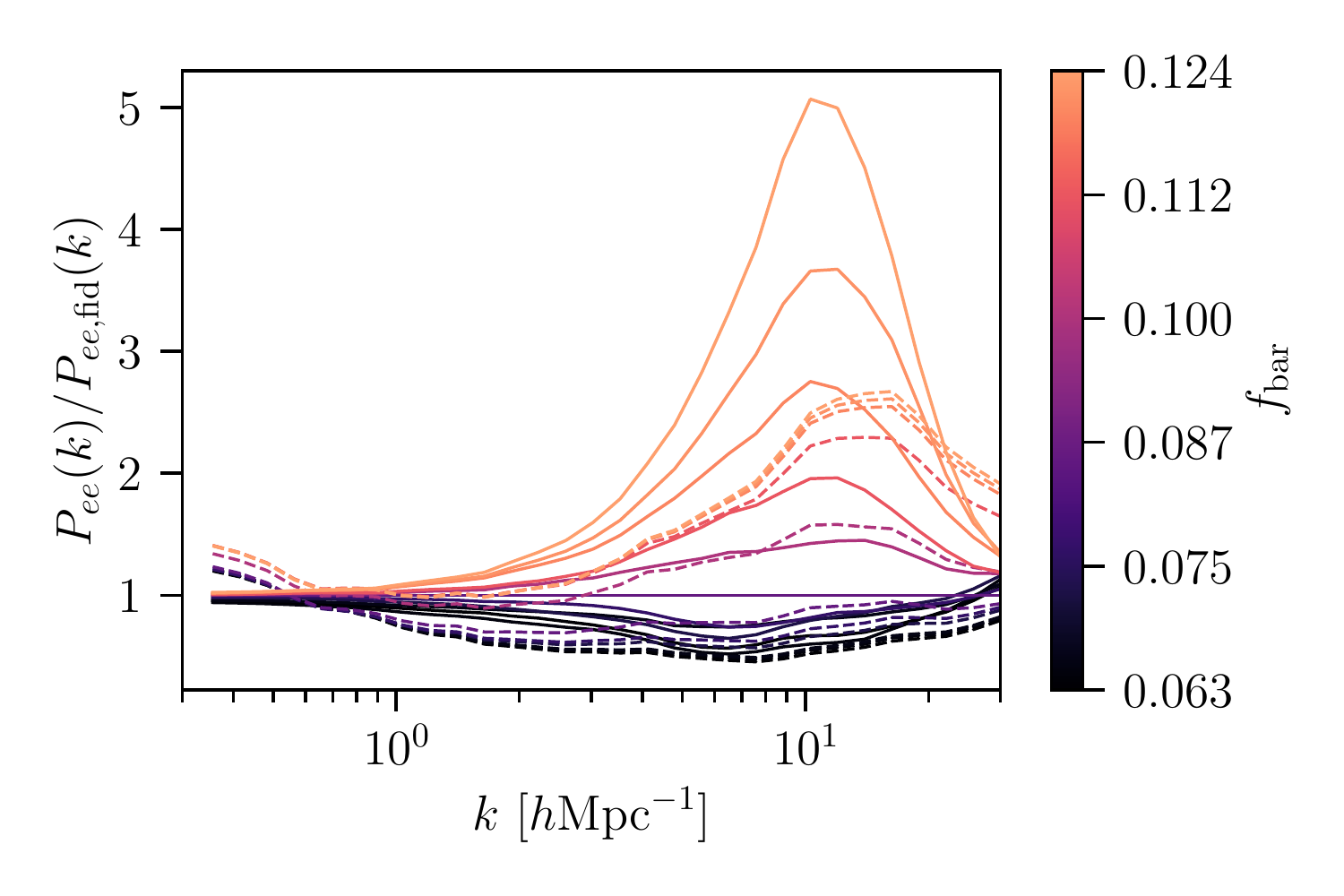}
 \caption{Comparison of the electron power spectra at redshift $z=0.0$ measured for the \textit{1P} set as a function of $\bar{f}_{\mathrm{bar}}$ and the emulated spectra from the Neural Network trained with the input parameter set $\boldsymbol{\vartheta} = (\Omega_{m}, \sigma_{8}, \bar{f}_{\mathrm{bar}})$ (same as the right hand panel of Fig.~\ref{fig:pkme_z0_no-cv_As-vs-fbar}). The power spectra are shown only for IllustrisTNG and for different SNe feedback strengths; for clarity, we have normalized all of them by the fiducial measurement in CAMELS, which is defined by $A_{\mathrm{SN}1} =1$.}
\label{fig:pkee_z0_cc=fbar}
\end{center}
\end{figure}

\begin{figure}
\begin{center}
\includegraphics[width=0.49\textwidth]{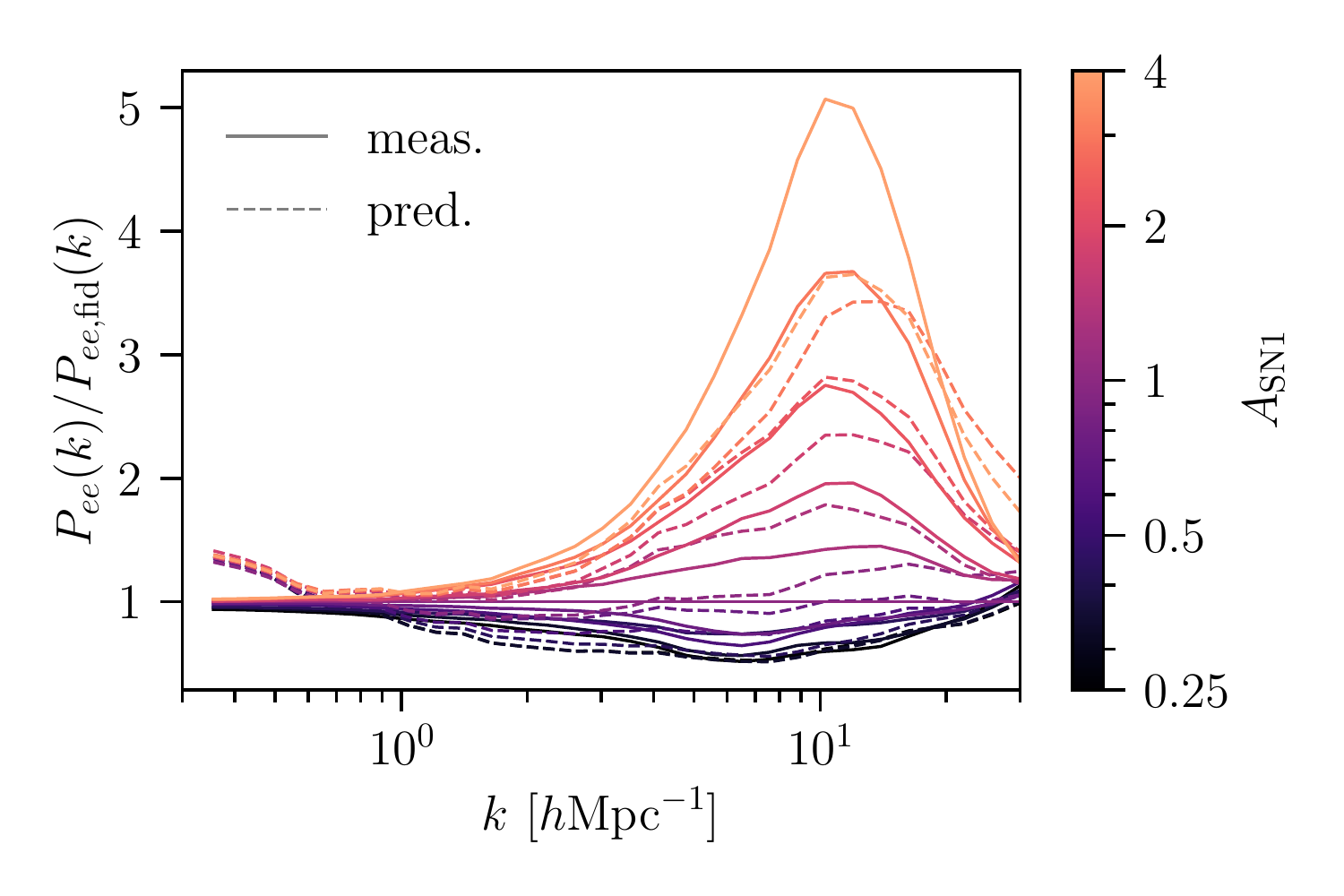}
 \caption{Comparison of the electron power spectra at redshift $z=0.0$ measured for the \textit{1P} set as a function of $A_{\mathrm{SN}1}$ and the emulated spectra from the Neural Network (same as the right hand panel of Fig.~\ref{fig:pkme_z0_no-cv_As-vs-fbar}) trained with the input parameter set $\boldsymbol{\vartheta} = (\Omega_{m}, \sigma_{8}, \bar{f}_{\mathrm{bar}}(r_{i}), i=1, 2, 3)$ as well as the cosmic variance parameter $\mathcal{M}_{h, \mathrm{box}}$. The power spectra are shown only for IllustrisTNG and for different SNe feedback strengths; for clarity, we have normalized all of them by the fiducial measurement in CAMELS, which is defined by $A_{\mathrm{SN}1} =1$.}
\label{fig:pkee_z0_cv_fbar-r}
\end{center}
\end{figure}

\section{Connection to observations} \label{ap:sec:P2obs}

Despite not propagating our results to cosmological observables, in the following we summarize the relations between the three-dimensional power spectra considered in this work and observable quantities.

Auto- and cross-correlations between the electron and pressure distributions with the matter and galaxy density are most readily measured through the spherical harmonic power spectrum, $C_{\ell}$. This quantity generally takes the form of a weighted integral over the three-dimensional power spectrum. Let $a, b$, denote two tracers of the LSS, which in this work we take to be the distribution of FRB dispersion measures, the thermal Sunyaev-Zel'dovich field, the spatial distribution of galaxies, as well as the weak lensing shear field, i.e. $a, b \in [\mathrm{FRB}, \mathrm{tSZ}, g, \gamma]$. Employing the Limber approximation \cite{Limber:1953, Kaiser:1992, Kaiser:1998}, we can write their spherical harmonic power spectrum as
\begin{equation}
\begin{aligned}
C_{\ell}^{ab}=\int \mathrm{d} z \; \frac{c}{H(z)} \; \frac{W^{a}\left(\chi(z)\right)W^{b}\left(\chi(z)\right)}{\chi^{2}(z)} \times \\
P_{ab}\left(k=\frac{\ell+\sfrac{1}{2}}{\chi(z)}, z\right).
\end{aligned}
\end{equation}
In the above equation, $c$ denotes the speed of light, $\chi(z)$ is the comoving distance to redshift $z$ and $P_{ab}(k, z)$ denotes the three-dimensional power spectrum between probes $a$ and $b$. The quantity $W^{a}\left(\chi(z)\right)$ is a probe-specific window function. In the following sections, we give a short derivation of the FRB window function as well as the explicit expressions for the window functions of all the probes considered in this analysis.

\subsection{Fast Radio Bursts} \label{sssec:frbs}

As discussed in Sec.~\ref{sec:intro}, Fast Radio Bursts are short pulses of radio emission, which exhibit frequency-dependent arrival times, i.e. they are dispersed due to interactions between the pulses and an an intervening ionized medium (for a review of FRBs, see e.g. Ref.~\cite{Petroff:2019}). These dispersion measures (DM) can be determined from FRB observations and provide a measure for the ionized electron density integrated along the line of sight. The total DM of an FRB located at source redshift $z_{s}$ is given by \cite{Petroff:2019}
\begin{equation}
\mathrm{DM} = \mathrm{DM}_{h} + \mathrm{DM}_{\mathrm{IGM}} + \mathrm{DM}_{\mathrm{Gal}},
\end{equation}
where $\mathrm{DM}_{h}$ is the contribution from the FRB host, $\mathrm{DM}_{\mathrm{IGM}}$ denotes the cosmological contribution due to the Inter-Galactic Medium (IGM) along the line-of-sight and $\mathrm{DM}_{\mathrm{Gal}}$ is due to the Galaxy. The information on the baryon distribution in the Universe is contained in $\mathrm{DM}_{\mathrm{IGM}}$. Following Ref.~\cite{Shirasaki:2017}, the IGM contribution can be written as 
\begin{equation}
\mathrm{DM}_{\mathrm{IGM}}(\boldsymbol{\theta}, z_{s}) = \int_{0}^{z_{s}} \mathrm{d}z \frac{c}{H(z)} \frac{n_{e}(\boldsymbol{\theta}, z) }{(1+z)^{2}}. 
\end{equation}
The quantity $n_{e}(\boldsymbol{\theta}, z)$ denotes the three-dimensional electron number density, which we can express as $n_{e}(\boldsymbol{\theta}, z) =n_{e}(z)(1+\delta_{e}(\boldsymbol{\theta}, z))$. This allows us to write \cite{Reischke:2020}
\begin{equation}
\mathrm{DM}_{\mathrm{IGM}}(\boldsymbol{\theta}, z_{s}) = \overline{\mathrm{DM}}_{\mathrm{IGM}}(z_{s}) + \Delta \mathrm{DM}_{\mathrm{IGM}}(\boldsymbol{\theta}, z_{s}),
\end{equation}
where the DM fluctuations induced by fluctuations in the LSS are given by
\begin{equation}
\Delta \mathrm{DM}_{\mathrm{IGM}}(\boldsymbol{\theta}, z_{s}) = \int_{0}^{z_{s}} \mathrm{d}z \frac{c}{H(z)} \frac{n_{e}(z)\delta_{e}(\boldsymbol{\theta}, z)}{(1+z)^{2}}. 
\end{equation}
Finally, for a population of FRBs with normalized redshift distribution $n_{\mathrm{FRB}}(z_{s})$ we obtain the DM fluctuation field as
\begin{equation}
\Delta \mathrm{DM}_{\mathrm{IGM}}(\boldsymbol{\theta}) = \int_{0}^{z_{\mathrm{max}}} \mathrm{d}z\; \frac{c}{H(z)} W_{\mathrm{FRB}}(\chi(z)) \delta_{e}(\boldsymbol{\theta}, z),
\label{eq:DM_source}
\end{equation}
where the window function for FRBs is defined as
\begin{equation}
W_{\mathrm{FRB}}(\chi(z))  = \frac{n_{e}(z)}{(1+z)^{2}} \int_{z}^{z_{\mathrm{max}}} \mathrm{d}z_{s}\; n_{\mathrm{FRB}}(z_{s}).
\end{equation}
From Eq.~\ref{eq:DM_source} we see that in the case of FRBs, we have that $P_{\mathrm{FRB} b}(k, z) \coloneqq P_{e b}(k, z)$, where $b \in [g, \gamma, \mathrm{FRB}, \mathrm{tSZ}]$.

\subsection{Thermal Sunyaev-Zel'dovich effect} \label{sssec:tSZ}

The thermal Sunyaev-Zel'dovich effect (tSZ) allows for the measurement of the Compton-$y$ parameter, which is given by \cite{Sunyaev:1972}
\begin{equation}
y(\boldsymbol{\theta}) = \frac{\sigma_{T}}{m_{e}c^{2}}\int \frac{\mathrm{d}\chi}{(1+z)} \; p(\chi \boldsymbol{\theta}).
\end{equation}
Here, $m_{e}$ denotes electron mass, $\sigma_{T}$ is the Thompson cross-section and $p(\chi \boldsymbol{\theta})$ denotes the three-dimensional electron pressure distribution. The window function associated to measurements of the Compton-$y$ parameter from tSZ surveys can be written as \cite{Koukoufilippas:2020}
\begin{equation}
W_{\mathrm{tSZ}}\left(\chi(z)\right) = \frac{\sigma_{T}}{m_{e}c^{2}}\frac{1}{(1+z)},
\end{equation}
which implies that the power spectrum associated to tSZ observations is given by $P_{\mathrm{tSZ} b}(k, z) = P_{p b}(k, z)$, where $b \in [g, \gamma, \mathrm{FRB}, \mathrm{tSZ}]$.

\subsection{Galaxy overdensity} \label{sssec:gc}

A galaxy sample with normalized redshift distribution $n_{g}(z)$ constrains the galaxy overdensity, integrated along the redshift direction $z$ with a window function given by:
\begin{equation}
W_{g}(\chi(z)) = \frac{H(z)}{c} n_{g}(z).
\end{equation}
Furthermore, the relevant three-dimensional power spectrum is $P_{g b}(k, z) \coloneqq P_{g b}(k, z)$, where $b \in [g, \gamma, \mathrm{FRB}, \mathrm{tSZ}]$.

\subsection{Weak gravitational lensing} \label{sssec:wl}

Weak gravitational lensing (WL) is sensitive to the integrated matter distribution between source galaxies and the observer, and the cosmic shear kernel $W^{\gamma}\left(\chi(z)\right)$ is given by
\begin{equation}
W_{\gamma}\left(\chi(z)\right) = \frac{3}{2} \frac{\Omega_{m} H^{2}_{0}}{c^{2}} \frac{\chi(z)}{a} \int_{\chi(z)}^{\chi_{h}} \mathrm{d} z' n_{\gamma}(z') \frac{\chi(z')-\chi(z)}{\chi(z')}.
\label{eq:gammawindow}
\end{equation}
In the above equation, $n_{\gamma}(z)$ denotes the normalized redshift distribution of source galaxies, $\chi_{h}$ is the comoving distance to the horizon and $a$ denotes the scale factor. As WL is sensitive to all gravitationally interacting matter in the Universe, we have that $P_{\gamma b}(k, z) \coloneqq P_{m b}(k, z)$, where $b \in [g, \gamma, \mathrm{FRB}, \mathrm{tSZ}]$.

\subsection{Kinematic Sunyaev-Zel'dovich effect} \label{sssec:kSZ}

As a final probe, we consider the kinematic Sunyaev-Zel'dovich effect (kSZ) \cite{Sunyaev:1972, Sunyaev:1980, Ostriker:1986}. The detection of this effect, however, requires us to abandon the general spherical harmonic framework described in the previous sections. The kSZ leads to a perturbation in the observed CMB temperature, $\Delta T_{\mathrm{kSZ}}$, given by
\begin{equation}
\frac{\Delta T_{\mathrm{kSZ}}}{T_{\mathrm{CMB}}}(\boldsymbol{\theta}) = -\sigma_{T}\int \frac{\mathrm{d}\chi}{(1+z)} \; e^{-\tau} n_{e}(\boldsymbol{\theta}, \chi) \boldsymbol{v}_{e}\boldsymbol{\theta},
\label{eq:kSZ}
\end{equation}
where $\tau$ is the optical depth to Thompson scattering and $\boldsymbol{v}_{e}$ denotes the peculiar velocity of the scattering electrons. As Eq.~\ref{eq:kSZ} depends on the electron velocity $\boldsymbol{v}_{e}$, any cross-correlation between the kSZ and other LSS probes will be suppressed, since electrons are equally likely to move towards or away from an observer. In order to probe the electron distribution using kSZ measurements, it is therefore necessary to consider higher order statistics such as the squared kSZ projected-fields estimator described in Refs.~\cite{Dore:2004, DeDeo:2005, Ferraro:2016, Hill:2016}. We refer the reader to these works for a more detailed description.

\bibliography{bibliography}{}

\end{document}